\documentclass[superscriptaddress,twocolumn,aps,pra,showpacs,floatfix]{revtex4}

\usepackage{dcolumn}
\usepackage{graphicx}
\usepackage{epsf}
\usepackage{amsmath}
\usepackage{bm}
\usepackage{times}

\def\half{{\textstyle{1\over 2}}}

\def\half{{\textstyle{\frac12}}}

\newcommand{\ii}{{\mathrm i}}

\newcommand{\addrHD}{Max--Planck--Institut f\"ur Kernphysik,
Postfach 103980, 69029 Heidelberg, Germany}

\newcommand{\addrROLLA}{Department of Physics,
Missouri University of Science and Technology,
Rolla MO65409-0640, USA}

\begin{document}

\sloppy

\title{Relativistic $\bm{(Z\alpha)^2}$---Corrections
and Leading Quantum Electrodynamic Corrections\\
to the Two--Photon Decay Rate of Ionic States}

\author{Benedikt J. Wundt}
\affiliation{\addrHD}

\author{Ulrich D. Jentschura}
\affiliation{\addrROLLA}

\begin{abstract}

We calculate the relativistic corrections of relative 
order $(Z\alpha)^2$ to the two-photon decay rate of higher
excited $S$ and $D$ states in ionic atomic systems, and we 
also evaluate the leading radiative corrections of relative
order $\alpha (Z\alpha)^2 \, \ln[(Z\alpha)^{-2}]$.
We thus complete the 
theory of the two-photon decay rates
up to relative order $\alpha^3 \, \ln(\alpha)$.
An approach inspired by nonrelativistic 
quantum electrodynamics is used.
We find that the corrections of relative order $(Z\alpha)^2$ 
to the two-photon decay are given by 
the zitterbewegung, the spin-orbit coupling and
by relativistic corrections to the electron mass,
and by quadrupole interactions. 
We show that all corrections are separately gauge-invariant 
with respect to a ``hybrid'' transformation from velocity to length gauge,
where the gauge transformation of the wave function is neglected.
The corrections are evaluated for the two-photon decay from
$2S$, $3S$, $3D$, and $4S$ states in one-electron (hydrogenlike)
systems, with $1S$ and $2S$ final states. 

\end{abstract}


\maketitle

%
%
\section{INTRODUCTION}
\label{intro}

Two-photon decay processes in hydrogen-like 
ions represent an intriguing physical 
phenomenon and are the subject of intense research. 
The metastability of the $2S$ level, 
which is limited only by two-photon decay,
makes it amenable to high-precision measurements.
Interestingly, though, the two-photon decay 
has never been studied within the 
so-called $Z\alpha$-expansion beyond leading 
order, that is, beyond the order of 
$\alpha^2(Z\alpha)^6$ for the decay width 
in units of the electron rest mass
(in this article, 
we use natural units, $\hbar = c = \epsilon_0 = 1$).

The first study of the two-photon decay 
rate $\Gamma$ of the 2$S$ state was carried out 
by G\"oppert-Mayer in 1931 \cite{GM1931}, 
and the well-known nonrelativistic result 
was derived,
\begin{equation}
\label{Gamma0}
\tau^{-1} = \Gamma_0 = 8.229\,352 \, Z^6 s^{-1} 
= 1.309\,742 \, Z^6 \, \mathrm{Hz} \,.
\end{equation}
This result has been verified 
experimentally~\cite{LiNoTo1965,MaSc1972,KoClNo1972}.

In the non-recoil limit,
the leading correction terms modifying this result
are given by a relativistic correction of relative 
order $(Z\alpha)^2$ and a radiative correction of 
relative order  $\alpha \, (Z\alpha)^2 \, \ln[(Z\alpha)^{-2}]$.
We can write the following expansion,
\begin{equation}
\label{gamma23}
\Gamma = \Gamma_0 \, 
\left[ 1 + \gamma_2 \, (Z\alpha)^2 + 
\gamma_3 \, \frac{\alpha}{\pi} \, (Z\alpha)^2 \, \ln[(Z\alpha)^{-2}] + \dots \right] \,,
\end{equation}
with coefficients $\gamma_2$ and $\gamma_3$ to be determined.

The next higher-order term not included in 
Eq.~\eqref{gamma23} is a nonlogarithmic 
radiative correction of order $\alpha\, (Z\alpha)^2$.
Equation~\eqref{gamma23} is complete up to order
$\alpha^3 \ln(\alpha)$.

The coefficient $\gamma_3$ is known for the $2S$-$1S$ 
transition~\cite{Je2004rad,KaIv1997c}, but it remains unknown 
for any other two-photon transition in a hydrogenlike ionic
system. The coefficient $\gamma_2$, which intuitively could be 
assumed to represent an easy computational task, 
has not yet been calculated for {\em any} two-photon transition,
to the best of our knowledge. 
We address both $\gamma_2$ and $\gamma_3$ in this paper.

The relativistic correction of 
relative order $(Z\alpha)^2$ actually involves
quite a large number of individual contributions:
(i) multipole (quadrupole radiation) correction, 
(ii) relativistic corrections to the 
electron's transition current, and
(iii) relativistic corrections to the Hamiltonian and 
to the bound-state energies of initial and final states, 
due to zitterbewegung, relativistic kinetic energy, and 
spin-orbit coupling. Each one of these contributions
entails a computationally demanding sum over 
virtual states, and an integration over the 
photon energy. We here calculate
the corrections one after the other and check
gauge invariance all along the way.
Finally, we obtain 
rigorous results for $\gamma_2$ and $\gamma_3$.

Our approach is inspired by Nonrelativistic 
Quantum Electrodynamics (NRQED), albeit in a 
restricted way: in a 
two-photon decay, the photon energies are bound by the 
energy difference of the initial and final states,
and therefore the problem of separating the energy scales
of the high-energy vertex terms does not arise.
However, the interaction Hamiltonian still
has to be expanded in the sense of NRQED, and 
we have the choice between 
two gauges which determine the form of the interaction 
Hamiltonian. Either the ``length'' (Yennie) or 
``velocity'' (Coulomb) gauge can be chosen. 
The final result should not depend on the gauge. 

In the Appendix of Ref.~\cite{GoDr1981}, 
the gauge invariance of the two-photon decay rate
was shown to hold within the fully relativistic 
formalism, within the class of fully relativistic 
gauge transformations given by Eq.~(A8) of Ref.~\cite{GoDr1981}.
The Power--Zienau gauge transformation~\cite{PoZi1959} as given in 
Eqs.~(18) and~(19) of Ref.~\cite{Pa2005} has a nontrivial dependence on the 
coordinates and allows us to express the 
QED interaction Hamiltonian exclusively in terms of
observable field strengths which in turn correspond to 
derivatives of the vector potential.
This transformation is most suitable for a nonrelativistic 
treatment, but due to the nontrivial dependence on the 
coordinates and due to problems related to the 
physical interpretation of non-gauge invariant 
quantities~\cite{Ko1978,ScBeBeSc1984,LaScSc1987}, a few subtleties arise.

After considerable discussion on this point
within the community~\cite{Ko1978,ScBeBeSc1984,LaScSc1987},
the conclusion has been reached 
that gauge transformations have to be considered 
very carefully in bound-state problems.
E.g., for the 
radiative corrections to the two-photon decay rate~\cite{Je2004rad}, 
the results are invariant under a ``hybrid''
gauge transformation \cite{ScBeBeSc1984},
where the interaction Hamiltonian is gauge 
transformed, but the gauge transformation of the
wave function is neglected.
I.e., although a gauge transformation normally 
entails a local, ``pointwise'' transformation of the wave function,
this whole transformation is flatly ignored, and the ``usual''
Schr\"{o}dinger eigenstates~\cite{LaLi1958vol3} are used for initial
and final states of the process under investigation. 
We show here that the relativistic 
corrections to the two-photon decay rate 
are invariant under such a transformation
(the gauge invariance of the leading logarithmic 
QED corrections was shown in Ref.~\cite{Je2004rad}).
In general, properties of atomic states which can be 
formulated using adiabatic $S$-matrix theory 
are invariant under this kind of hybrid gauge transformation,
whereas in time-dependent problems, the choice of gauge 
has to be taken into account even more 
carefully~\cite{Ko1978,ScBeBeSc1984,LaScSc1987}.
In the latter case,
the gauge transformation of the wave function cannot 
be ignored.

When generalizing the results to higher excited initial and final states, one
has to overcome a few subtle difficulties, because one has to separate the
$3S$-$1S$ double-dipole ($E1E1$) two-photon decay from the cascade
$3S$-$2P$-$1S$. The $2P$ state appears both as a virtual state for the
two-photon decay process as well as an intermediate state for the cascade
process. In the two-photon decay rate,
when regarded as differential with respect to the photon
energy, the presence of the $2P$ state causes a 
(quadratic) singularity. Because we are
interested in the total decay rate, we have to integrate over this singularity,
which is quadratic and thus {\em a priori} 
not integrable. Removing the $2P$ state from the
sum over virtual intermediate states leads to gauge-dependent
results~\cite{Fl1984,CrTaSaCh1986,FlScMi1988,ChSu2008,Je2008}. In order to
separate the cascade contribution from the two-photon correction for the
two-photon decay, one has to use a special integration prescription detailed in
Refs.~\cite{Je2007,Je2008,JeSu2008,Je2009}; the prescription constitutes a
generalization of the principal-value integration to quadratic singularities.
Here, we extend the relativistic calculations for two-photon decays to highly
excited initial states using this formalism.

We organize the paper as follows. In Sec.~\ref{Theo}, we explain the
theoretical methods used in our approach. 
In Sec.~\ref{genproof}, we consider all
the corrections separated by their physical origin for the $2S$-$1S$ transition
and show explicitly that each contribution is gauge invariant. In
Sec.~\ref{numerics}, we present numerical 
results for the $2S$-$1S$ transition and also for 
transitions from higher excited
states, and we discuss the separation of the cascade
contribution from the coherent two-photon correction to the decay rate.
Results for the QED radiative corrections of logarithmic
order are presented in Sec.~\ref{log}.
Conclusions are drawn in Sec.~\ref{Conclusions}. 
As already mentioned, natural units $\hbar =
\epsilon_0 = c=1$ are used throughout this paper.

%
%
\section{THEORETICAL BACKGROUND}
\label{Theo}

The two-photon decay rate is given as the imaginary part of the two-loop
self-energy correction \cite{BaSu1978} which can be derived using
nonrelativistic Quantum Electrodynamics (NRQED) \cite{Pa2001}. A detailed
derivation of the nonrelativistic two-photon decay rate, valid for all
transitions including those involving highly excited states, is contained in
previous works \cite{Je2004rad,Je2007,Je2008,Je2009}, and there is no need to
reproduce it here. 

We recall that in velocity (Coulomb) gauge, the interaction Hamiltonian 
for the interaction of the electron with the 
quantized radiation field is given as
\begin{equation}
H_I = - \frac{e}{2m} \left( \vec{p} \cdot \vec{A} + 
\vec{A} \cdot \vec{p} \right) + \frac{e^2 \vec{A}^2}{2m} \,,
\end{equation}
where $\vec{p}$ is the electron momentum,
$\vec{A}$ is the vector potential, and $m$ is the electron mass.
This interaction leads to the following expression for the nonrelativistic 
decay rate,
\begin{align}
\Gamma^\xi \!\! &= \! \frac{4 \alpha^2}{9 \pi m^4} 
\mathrm{Re} \!\!\!\!\!\! \int \limits_0^{E_i-E_f} \!\!\!\!\!\!\! 
d \omega_1 \omega_1 \omega_2 \! \biggl( \!\! 
\left< \! \Phi_f \! \left| p^i \frac{1}{H \!-\!E_f\!
-\!\omega_1\!+\!\mathrm{i} \epsilon}  p^j \right| \!\Phi_i\! \right> \nonumber \\
& \quad + \! \left< \! \Phi_f \! \left| p^i \frac{1}{H \!-\!E_i
\!+\!\omega_1\!+\!\mathrm{i} \epsilon} p^j 
\right| \!\Phi_i\! \right> \! \biggr)^2 \, \label{gamvg} \,.
\end{align}
Here, ``Re'' denotes the real part,
and the limit $\epsilon \to 0$ is taken after all integrations 
have been performed. The summation convention 
is used throughout this article. The superscript $\xi$ denotes
the velocity-gauge form of the expression.

For length (Yennie) gauge, 
the (leading) interaction Hamiltonian takes the simple form
\begin{equation}
H_I = - e \vec{E} \cdot \vec{r} \,.
\end{equation}
If this Hamiltonian is used,
we obtain for the nonrelativistic expression
\begin{align}
\Gamma^\zeta \!\! &= \! \frac{4 \alpha^2}{9 \pi} 
\mathrm{Re} \!\!\!\!\!\! \int \limits_0^{E_i-E_f} \!\!\!\!\!\! 
d \omega_1 \omega_1^3 \omega_2^3 \biggl( \!\! 
\left< \! \Phi_f \! \left| r^i \frac{1}{H \!-\!E_f \!-\!\omega_1\!+\!\mathrm{i} \epsilon} r^j 
\right| \!\Phi_i\! \right> \nonumber \\
& \quad + \! \left< \! \Phi_f \! \left| r^i 
\frac{1}{H \!-\!E_i \!+\!\omega_1\!+\!\mathrm{i} \epsilon} 
r^j \right| \!\Phi_i\! \right> \biggr)^2 \,. \label{gamlg}
\end{align}
Using the relation \cite{Ko1978,BaFoQu1977}
\begin{align}
& \left< \! \Phi_f \! \left| p^i \frac{1}{H \!-\!E_f\!
-\!\omega_1}  p^j \right| \!\Phi_i\! \right> \! + \! 
\left< \! \Phi_f \! \left| p^i \frac{1}{H \!-\!E_i\!
+\!\omega_1}  p^j \right| \!\Phi_i\! \right> \nonumber \\
&= -m^2 \omega_1 \omega_2 \left( \left< \! \Phi_f \! \left| r^i 
\frac{1}{H \!-\!E_f \!-\!\omega_1} r^j 
\right| \!\Phi_i\! \right> \right. \nonumber \\
& \qquad \left. + \! \left< \! \Phi_f \! \left| r^i 
\frac{1}{H \!-\!E_i \!+\!\omega_1} 
r^j \right| \!\Phi_i\! \right> \right) \,, \label{gaugenr}
\end{align}
the equivalence of these two expressions can be shown. 
Note that this is only valid if a complete spectrum is 
used for the representation of the 
propagator.

For a fully relativistic calculation of the effect,
we would have to use the Dirac Hamiltonian
$H_D$ in the propagators instead of the Schr\"odinger Hamiltonian $H$, and also
the interaction Hamiltonian and the wavefunction would have to be changed
accordingly. However, as we want to work nonrelativistically, we transform the
fully relativistic Dirac Hamiltonian and its interaction Hamiltonian into
effective nonrelativistic operators. This can be achieved by using a
Foldy-Wouthuysen transformation~\cite{FoWu1950}, which identifies the
nonrelativistic Hamiltonian as the leading term, and thus leads to a systematic way of
expressing the relativistic corrections. Furthermore, it allows us 
to express the
relativistic corrections to the electron's transition current within the
$Z\alpha$ expansion.

Alternatively, one can resort to the literature~\cite{CaLe1986},
where the corrections to the Schr\"{o}dinger Hamiltonian 
have been tabulated. For the non-interacting part, this 
procedure leads to the well-known 
corrections to the Schr\"odinger Hamiltonian $H$,
\begin{equation}
\label{H4}
\begin{split}
H &\rightarrow H + \delta H \,, \\
H &= \frac{p^2}{2m} + \frac{Z\alpha}{r} \,, \\
\delta H &= \frac{\pi Z \alpha}{2m} \delta^3 (r) 
+ \frac{\vec{L} \cdot \vec{\sigma}}{4m^2 r^3} - \frac{p^4}{8m^3} \,.
\end{split}
\end{equation}
The Darwin term proportional to the Dirac $\delta$ originates from 
the zitterbewegung of the electron. The next term is 
the spin-orbit coupling, and the last is the correction due 
to the relativistic kinetic energy. The 
relativistic corrections to the reference state wavefunction 
and to its energy thus read as follows,
\begin{align}
E &\rightarrow E + \delta E
= E + \left< \Phi \left| \delta H \right| \Phi \right> \,, 
\\[2ex]
\left| \Phi \right> &\rightarrow \left| \Phi \right> 
+ \left| \delta \Phi \right> 
= \left| \Phi \right> + \left( \frac{1}{E-H} \right)' 
\delta H \left| \Phi \right> \,.
\end{align}
The transition current of the electron can be derived by 
acting with the Foldy--Wouthuysen transformation 
on a Dirac Hamiltonian which 
is coupled to an electromagnetic vector potential.
The velocity-gauge result for the 
interaction Hamiltonian thus is (see Refs.~\cite{Pa2005,JeCzPa2005})
\begin{align}
\label{hivg}
& H_{\mathrm{int}} = -\frac{e \vec{A} \cdot \vec{p}}{m} 
-\frac{e}{2m} \left( \vec{\sigma} \times \vec{\nabla} \right) 
\cdot \vec{A} +
\frac{e}{2m^3} \left( \vec{A} \cdot \vec{p} \right) \, \vec{p}^{\,2}
\nonumber\\
& \; -\frac{e}{4m^2} \left(\vec{\sigma} \times \vec{p} \right) 
\! \cdot \!\! \frac{\partial \vec{A}}{\partial t} 
-\frac{e}{4m^2} \left( \vec{\sigma} \times \vec{\nabla}V \! \right) 
\! \cdot \! \vec{A} 
\equiv -e \vec{J} \cdot \vec{A} \,.
\end{align}
We remember that the photon emission is characterized by the 
creation part of the electromagnetic vector potential operator,
which carries a dependence of $\exp(-\ii \, \vec k \cdot \vec r)$.
The transition current $\vec{J}$ can thus be written as
\begin{align}
\label{delJvg}
& J^i = \; \frac{p^i}{m} + \delta J^i =
\; \frac{p^i}{m} \left(1-\mathrm{i} \, \vec{k} \cdot \vec{r} 
-\half \, (\vec{k} \cdot \vec{r})^2 \right) 
- \frac{p^i \, \vec{p}^{\,2}}{2m^3} 
\nonumber\\[2ex]
& -\frac{1}{2m^2} \frac{Z \alpha}{r^3} \left( \vec{r} 
\times \vec{\sigma} \right)^i 
- \frac{\mathrm{i}}{2m} \left( \vec{\sigma} \times \vec{k} \right)^i 
\left( 1 - \mathrm{i} \, \vec{k} \cdot \vec{r} \right)  \,.
\end{align}
As we are considering a two-photon effect, 
contributions from seagull terms also
have to be taken into account (here, two photons emerge
from the same vertex). Terms proportional 
to $A^2$ are included in the seagull Hamiltonian 
which is given by
\begin{equation}\label{hsea}
H_{\mathrm{sea}} = \frac{e^2 \vec{A}^2}{2m} 
-\frac{e^2}{2m^3} \left( \vec{A} \cdot \vec{p} \right)^2 
- \frac{e^2}{4m^3} \vec{A}^2 \vec{p}^{\,2} \,.
\end{equation}
Expanding in powers of $(Z\alpha)$ and extracting the 
photon creation part, 
we obtain the seagull correction in relative order $(Z\alpha)^2$,
\begin{equation}
\label{seagull}
\delta S^{ij} = 
-\frac{1}{2m} (\vec{k} \cdot \vec{r})^2 \delta^{ij}
-\frac{p^i p^j}{2 m^3} 
- \frac{p^2}{4 m^3} \delta^{ij} \,,
\end{equation}
written in such a way that it multiplies the (creation part of the)
photon fields $A^i \, A^j$.

The interaction Hamiltonian in length gauge,
including relativistic and multipole corrections,
can be obtained by employing two consecutive 
Power-Zienau transformations \cite{PoZi1959} after the 
Foldy-Wouthuysen transformation. This has been shown in 
Ref.~\cite{Pa2004}. The interaction Hamiltonian 
in length gauge thus reads
\begin{align}\label{hilg}
H_{\mathrm{int}} &= - e \vec{r} \cdot \vec{E} 
-\frac{e}{2m} \left( \vec{L} + \vec{\sigma} \right) 
\cdot \vec{B} - \frac{e}{2} r^i r^j E^i_{,j} 
\nonumber\\
& \quad - \frac{e}{6m} \left( L^i \, r^j + r^j \, L^i \right) 
B^i_{,j} - \frac{e}{2m} \sigma^i r^j B^i_{,j} 
\nonumber\\
& \quad -\frac{e}{6} r^i r^j r^k E^i_{,jk} 
+ \frac{e}{4 m} \vec{\sigma} 
\left( \dot{\vec{E}} \times \vec{r} \right) \,.
\end{align}
Here, the subscript separated by commas denotes the spatial 
derivatives with respect to the indicated Cartesian coordinates, 
evaluated at the origin~\cite{Pa2004}, which is defined 
to be the location of the ionic nucleus.
This corresponds to a length-gauge transition current
\begin{align}
\label{delIlg} 
& I^i \equiv  r^i + \delta I^i 
= r^i \left(1 - \tfrac{\mathrm{i}}{2} \vec{k} \cdot \vec{r} 
- \tfrac{1}{6} (\vec{k} \cdot \vec{r})^2 \right) 
\nonumber \\
& 
+ \frac{\mathrm{i}\omega}{4m} \left( \vec{\sigma} \times \vec{r} \right)^i 
+ \frac{1}{2m\omega} \left( \vec{\sigma} \times \vec{k} \right)^i 
\left( 1 - \mathrm{i}\vec{k} \cdot \vec{r} \right)  
\nonumber\\
& + \frac{1}{2m\omega} ( \vec{L} \times \vec{k} )^i 
- \frac{\ii}{6m\omega} 
\left\{ ( \vec{L} \times \vec{k} )^i, \vec{k}\cdot \vec{r} \right\} \,,
\end{align}
where $\{ A, B \} = A\, B + B\, A$ is the anticommutator,
and we examine the emission of a photon with four-vector $(\omega, \vec k)$.
We are now in the position to discuss how the 
corrections to the decay rate can be
determined from the transition currents
in the two different gauges. We start with the
velocity gauge.

%
%
\subsection{Velocity Gauge}

The nonrelativistic two-photon decay rate in velocity 
gauge [see Eq.~\eqref{gamvg}] can be written as
\begin{equation}\label{gam0vg}
\Gamma^\xi = \frac{4\alpha^2}{9\pi} \!\!\!\!\! 
\int \limits_{0}^{E_{\Phi_i}-E_{\Phi_f}} \!\!\!\!\!\! 
d \omega_1 \, \omega_1 \, \omega_2 \, \xi^2 \,,
\end{equation}
where the superscript $\xi$ denotes the velocity-gauge expression.
Here, due to energy conservation,
$\omega_2=E_{\Phi_i}-E_{\Phi_f}-\omega_1$, and 
\begin{subequations}
\begin{align}
\xi=& \; \xi_1 +\xi_2 \,, \\
\xi_1 =& \; \left< \Phi_f \left| \frac{p^i}{m} 
\frac{1}{H-E_{\Phi_i} + \omega_1} \frac{p^j}{m} 
\right| \Phi_i \right> \,, \\
\xi_2 =& \; \left< \Phi_f \left| \frac{p^i}{m} 
\frac{1}{H-E_{\Phi_f} - \omega_1} \frac{p^j}{m} 
\right| \Phi_i \right> \,.
\end{align}
\end{subequations}
For the gauge invariance of this nonrelativistic expression,
see Eq.~\eqref{gaugenr}. We only remark that 
the statement of gauge invariance can be brought 
into the compact from
\begin{equation}\label{gaugesim}
\xi = - \omega_1 \, \omega_2 \, \zeta \,,
\end{equation}
where $\zeta$ is defined in Eq.~\eqref{defzeta} below.
Here and in the following, we suppress the superscripts
$ij$ of the $\xi$ and $\zeta$ tensors in order to ensure the 
compactness of the notation, and we imply that 
$\xi^2 \equiv \xi^{ij} \, \xi^{ij}$ (the indices
$i$ and $j$ are summed over), and that 
$\xi \delta\xi \equiv \xi^{ij} \, \delta\xi^{ij}$.
We define $\delta\xi$ to denote the sum of the corrections
due to all the previously discussed perturbations (Hamiltonian, 
energy, and current) and
express the first-order relativistic 
correction $\delta \Gamma$ to the decay rate
as (see Ref.~\cite{Je2004rad})
\begin{equation}\label{delgamvg}
\delta \Gamma = 2\frac{4\alpha^2}{9\pi} \!\!\!\!\!\!\!\! 
\int \limits_{0}^{E_{\Phi_i}-E_{\Phi_f}} \!\!\!\!\!\!\!\!\! 
d \omega_1 \, \omega_1 \, \omega_2 \, \xi \, \delta \xi 
+ \frac{4\alpha^2}{9\pi} \delta \omega_{\rm max}  \!\!\!\!\!\!\!\!\!\! 
\int \limits_{0}^{E_{\Phi_i}-E_{\Phi_f}} \!\!\!\!\!\!\!\!\! 
d \omega_1 \, \omega_1 \, \xi^2 \,.
\end{equation}
The correction 
$\delta \omega_{\rm max} = \delta E_{\Phi_i} - \delta E_{\Phi_f}$ 
is necessary to ensure that the perturbed energy conservation 
condition is fulfilled:
\begin{subequations}
\begin{align}
\omega_1 + \omega_2 =& \; E_{\Phi_i}- E_{\Phi_f} + \delta\omega_{\rm max} 
\,,
\\[2ex]
\label{omegamax}
\delta\omega_{\rm max} =& \;
\langle \Phi_i | \delta H | \Phi_i \rangle - 
\langle \Phi_f | \delta H | \Phi_f \rangle  \,.
\end{align}
\end{subequations}
so that the frequencies of the two quanta add up to the perturbed 
transition frequency. However, due to the presence 
of the seagull terms, further corrections have to 
be taken into account. 

After some algebra, we see that $\delta \xi$ can
be expressed as the sum of fifteen terms that account for all the 
relativistic and multipole perturbations, 
\begin{equation}
\delta \xi = \sum_{k=1}^{15} \delta \xi_k \,.
\end{equation}
The perturbations of the energies of the initial and final 
states lead to the following terms,
\begin{subequations}
\begin{align}
\delta \xi_1 \! &= \! \left< \! \Phi_f \! \left| 
\frac{p^i}{m} \! \left( \! 
\frac{1}{H \! - \! E_{\Phi_i} \! + \! \omega_1} 
\! \right)^2 \!\! \frac{p^j}{m} 
\right| \! \Phi_i \! \right> \! \left< \Phi_i \! \left| 
\delta H \right| \! \Phi_i \right>, \label{cde1vg} \\
\delta \xi_2 \! &=\! \left< \Phi_f \! \left| 
\delta H \right| \!\Phi_f \right> \! 
\left< \!\! \Phi_f \! \left| \! \frac{p^i}{m} \! \left( \! 
\frac{1}{H \! - \! E_{\Phi_f} \! - \! \omega_1} 
\! \right)^2 \!\! \frac{p^j}{m} 
\right| \!\Phi_i \!\! \right>. \label{cde2vg}
\end{align}
The perturbations to 
the initial and final-state wavefunctions
lead to the following four effects,
\begin{align}
\delta \xi_3 \! &=\! \left< \! \Phi_f \! \left| 
\frac{p^i}{m} 
\frac{1}{H \! - \! E_{\Phi_i} \! + \! \omega_1} 
\frac{p^j}{m} \! \left(\! 
\frac{1}{E_{\Phi_i} \! - \! H} \! \right)' 
\!\! \delta H \right| \!\Phi_i \! \right>  
, \label{cwf1vg} \\
\delta \xi_4 \! &=\! \left< \! \Phi_f \! \left| 
\frac{p^i}{m} \frac{1}{H \! - \! E_{\Phi_f} \!\! - \! \omega_1} 
\frac{p^j}{m} \! \left(\! 
\frac{1}{E_{\Phi_i} \! - \! H} \! \right)' 
\!\! \delta H \right| \! \Phi_i \! \right>  ,\label{cwf2vg} \\
\delta \xi_5 \! &=\! \left< \! \Phi_f \! \left| 
\delta H \!\! \left(\! \frac{1}{E_{\Phi_f} \! - \! H} 
\! \right)' \!\! \frac{p^i}{m} 
\frac{1}{H \! - \! E_{\Phi_i} \! + \! \omega_1} 
\frac{p^j}{m} \right| \! \Phi_i \! \right> , \label{cwf3vg} \\
\delta \xi_6 \!&=\! \left< \! \Phi_f \! \left| 
\delta H  \!\! \left(\! \frac{1}{E_{\Phi_f} \! - \! H} 
\! \right)' \! \frac{p^i}{m} \frac{1}{H \! - \! E_{\Phi_f} 
\!\! - \! \omega_1} \frac{p^j}{m} 
\right| \! \Phi_i \! \right> .\label{cwf4vg}
\end{align}
The perturbation incurred by the Hamiltonian 
leads to two terms (observe the different denominators),
\begin{align}
\delta \xi_7 \! &= \!-\! 
\left< \!\Phi_f\! \left| \frac{p^i}{m} 
\frac{1}{H \! - \! E_{\Phi_i} \! + \! \omega_1} 
\delta H \frac{1}{H \! - \! E_{\Phi_i} \! + \! \omega_1} 
\frac{p^j}{m} \right| \! \Phi_i \! \right> 
, \label{cdh1vg} \\
\delta \xi_8 \! &= \! - \! 
\left< \!\Phi_f\! \left| \frac{p^i}{m} 
\frac{1}{H \! - \! E_{\Phi_f}  \! - \! \omega_1} 
\delta H \frac{1}{H \! - \! E_{\Phi_f} \! - \! \omega_1} 
\frac{p^j}{m} \right| \!\Phi_i\! \right>\,.\label{cdh2vg}
\end{align}
The correction to the electron's transition current 
can affect both the initial and the final states,
and this gives rise to a total of four terms,
\begin{align}
\delta \xi_{9} &= \left< \!\Phi_f\! \left| \frac{p^i}{m} 
\frac{1}{H-E_{\Phi_i} +\omega_1} \delta J^j 
\right| \!\Phi_i\! \right> \,, 
\label{cdj1vg} \\
\delta \xi_{10} &= \left< \!\Phi_f\! \left| \frac{p^i}{m} 
\frac{1}{H-E_{\Phi_f}-\omega_1} \delta J^j 
\right| \!\Phi_i\! \right> \,, 
\label{cdj2vg}\\
\delta \xi_{11} &= \left< \!\Phi_f\! \left| \delta J^i 
\frac{1}{H-E_{\Phi_i} +\omega_1} \frac{p^j}{m} 
\right| \!\Phi_i\! \right>\,, 
\label{cdj3vg}\\
\delta \xi_{12} &= \left< \!\Phi_f\! \left| \delta J^i 
\frac{1}{H-E_{\Phi_f}-\omega_1} \frac{p^j}{m} 
\right| \!\Phi_i\! \right> \,.
\label{cdj4vg}
\end{align}
The seagull Hamiltonian acting on the 
unperturbed wavefunctions leads to
\begin{equation}\label{csea}
\delta \xi_{13} = -\left< \!\Phi_f\! \left| 
\delta S^{ij} \right| \!\Phi_i\! \right>  \,.
\end{equation}
The minus sign originates because we have written 
all matrix elements (second-order perturbations) in the 
``$1/(H-E)$'' form, which corresponds to a negative 
second-order energy perturbation. In order to be consistent,
we have to use the negative higher-order seagull Hamiltonian, 
which is applied in first-order perturbation theory.
Finally, we have the seagull terms 
which were already present in Ref.~\cite{Je2004rad} 
which account for the emission of two photons from 
the perturbed initial state or to the perturbed 
final state. They are given as
\begin{align}
\delta \xi_{14} &= -\frac{1}{m} \left< \!\Phi_f\! \left| 
\left(\frac{1}{E_{\Phi_i} -H} \right)' \delta H \right| 
\!\Phi_i\! \right> \, \delta^{ij} \,,\label{cseaula} \\
\delta \xi_{15} &= -\frac{1}{m} \left< \!\Phi_f\! \left| 
\delta H \left(\frac{1}{E_{\Phi_f} -H} \right)' 
\right| \!\Phi_i\! \right> \, \delta^{ij} \,,\label{cseaulb}
\end{align}
\end{subequations}
where we invoke second-order perturbation theory 
with the leading seagull term $e^2 \vec A^2/(2 m)$.
Using a complete basis-set of hydrogen eigenfunctions 
and their orthonormality relations, we can show that
\begin{equation}
\delta \xi_{14} + \delta \xi_{15} =0 \,.
\end{equation}
The reason is that both $\delta \xi_{14}$ and 
$\delta \xi_{15}$ are proportional to the non-diagonal
matrix element $\left< \!\Phi_f\! \left| \delta H \right| 
\!\Phi_i\! \right>$, but with opposite prefactors.

%
%
\subsection{Length Gauge}

The nonrelativistic, length gauge expression in 
Eq.~\eqref{gamlg} can be written as
\begin{equation}\label{gam0lg}
\Gamma^\zeta = \frac{4\alpha^2}{9\pi} \!\!\!\!\! 
\int \limits_{0}^{E_{\Phi_i}-E_{\Phi_f}} \!\!\!\!\!\! 
d \omega_1 \omega_1^3 \omega_2^3 \zeta^2 \,,
\end{equation}
where the superscript $\zeta$ denotes the length-gauge 
expression. Here, $\omega_2$ is defined as in Eq.~\eqref{gam0vg}, and 
\begin{subequations}
\label{defzeta}
\begin{align}
\zeta =& \zeta_1 +\zeta_2 \,, \\
\zeta_1 =& \left< \Phi_f \left| 
r^i \, \frac{1}{H-E_{\Phi_i} + \omega_1} \,
r^j \right| \Phi_i \right> \,, \\
\zeta_2 =& \left< \Phi_f \left| r^i \,
\frac{1}{H-E_{\Phi_f} - \omega_1} \, r^j 
\right| \Phi_i \right> \,.
\end{align}
\end{subequations}
Following the same procedure as for the 
velocity gauge expression, we can write the 
first-order correction to the two-photon 
decay rate in length gauge,
\begin{equation}\label{delgamlg}
\delta \Gamma^\zeta = 2\frac{4\alpha^2}{9\pi} \!\!\!\!\!\!\!\! 
\int \limits_{0}^{E_{\Phi_i}-E_{\Phi_f}} \!\!\!\!\!\!\!\!\! 
d \omega_1 \omega_1^3 \, \omega_2^3 \, \zeta \, \delta \zeta 
+ 3\frac{4\alpha^2}{9\pi} \delta \omega_{\rm max} \!\!\!\!\!\!\!\!\!\! 
\int \limits_{0}^{E_{\Phi_i}-E_{\Phi_f}} \!\!\!\!\!\!\!\!\! 
d \omega_1 \omega_1^3 \, \omega_2^2 \, \zeta^2 \,,
\end{equation}
where again $\delta \zeta$ denotes the sum 
of all the correction terms incurred by the relativistic 
perturbations of the Hamiltonian, and of the energies of the 
initial and final states, and of the length-gauge current.
Indeed, in the length gauge, the correction $\delta \zeta$ 
contains only twelve as opposed to fifteen terms,
\begin{equation}
\delta \zeta = \sum_{k=1}^{12} \delta \zeta_k \,.
\end{equation}
The energies of the initial and final states are perturbed and 
this gives rise to the first two correction terms,
\begin{subequations}
\begin{align}
\delta \zeta_1 \! &= \! 
\left< \! \Phi_f \! \left| r^i \! \left( \! 
\frac{1}{H \! - \! E_{\Phi_i} \! + \! \omega_1} 
\! \right)^2 \!\! r^j \right| \! \Phi_i \! \right> \! 
\left< \Phi_i \! \left| \delta H 
\right| \! \Phi_i \right>, \label{cde1lg} \\
\delta \zeta_2 \! &=\! \left< \Phi_f \! \left| 
\delta H \right| \!\Phi_f \right> \! 
\left< \!\! \Phi_f \! \left| r^i \! \left( \! 
\frac{1}{H \! - \! E_{\Phi_f} \! - \! \omega_1} 
\! \right)^2 \!\! r^j \right| 
\!\Phi_i \!\! \right>. \label{cde2lg}
\end{align}
In complete analogy to Eqs.~\eqref{cwf1vg}---\eqref{cwf4vg},
the perturbations to the 
initial and final state wavefunctions 
are accounted for by the following four terms,
\begin{align}
\delta \zeta_3 \! &=\! \left< \! \Phi_f \! \left| 
r^i \frac{1}{H \! - \! E_{\Phi_i} \! + \! \omega_1} 
r^j \! \left(\! \frac{1}{E_{\Phi_i} \! - \! H} 
\! \right)' \!\! \delta H 
\right| \!\Phi_i \! \right> , \label{perwf1} \\
\delta \zeta_4 \! &=\! \left< \! \Phi_f \! \left| 
r^i \frac{1}{H \! - \! E_{\Phi_f} \! - \! \omega_1} 
r^j \! \left(\! \frac{1}{E_{\Phi_i} \! - \! H} 
\! \right)' \!\! \delta H 
\right| \! \Phi_i \! \right>  ,\label{perwf2} \\
\delta \zeta_5 \! &=\! \left< \! \Phi_f \! \left| 
\delta H \!\! \left(\! \frac{1}{E_{\Phi_f} \! - \! H} 
\! \right)' \! r^i 
\frac{1}{H \! - \! E_{\Phi_i} \! + \! \omega_1} 
r^j \right| \! \Phi_i \! \right> ,\label{perwf3} \\
\delta \zeta_6 \!&=\! \left< \! \Phi_f \! \left| \delta H  
\!\! \left(\! \frac{1}{E_{\Phi_f} \! - \! H} \! \right)' 
\! r^i \frac{1}{H \! - \! E_{\Phi_f} \! - \! \omega_1} 
r^j \right| \! \Phi_i \! \right> .\label{perwf4}
\end{align}
Furthermore, the corrections from 
the perturbed Hamiltonian give rise to two terms,
\begin{align}
\delta \zeta_7 \! &= \!-\! \left< \!\Phi_f\! \left| 
r^i \frac{1}{H \! - \! E_{\Phi_i} \! + \! \omega_1} 
\delta H \frac{1}{H \! - \! E_{\Phi_i} \! + \! \omega_1} 
r^j \right| \! \Phi_i \! \right> , \label{cdh1lg} \\
\delta \zeta_8 \! &= \! - \! \left< \!\Phi_f\! \left| 
r^i \frac{1}{H \! - \! E_{\Phi_f}  \! - \! \omega_1} 
\delta H \frac{1}{H \! - \! E_{\Phi_f} \! - \! \omega_1} 
r^j \right| \!\Phi_i\! \right>\,.\label{cdh2lg}
\end{align}
The length-gauge correction to the current 
$\delta I$ gives rise to four more terms,
\begin{align}
\delta \zeta_{9} &= \left< \!\Phi_f\! \left| 
r^i \frac{1}{H-E_{\Phi_i} +\omega_1} \delta I^j 
\right| \!\Phi_i\! \right> \,, 
\label{cdi1lg}\\
\delta \zeta_{10} &= \left< \!\Phi_f\! \left| r^i 
\frac{1}{H-E_{\Phi_f}-\omega_1} \delta I^j 
\right| \!\Phi_i\! \right> \,, 
\label{cdi2lg}\\
\delta \zeta_{11} &= \left< \!\Phi_f\! \left| 
\delta I^i \frac{1}{H-E_{\Phi_i} +\omega_1} r^j 
\right| \!\Phi_i\! \right>\,, 
\label{cdi3lg}\\
\delta \zeta_{12} &= \left< \!\Phi_f\! \left| 
\delta I^i \frac{1}{H-E_{\Phi_f}-\omega_1} 
r^j \right| \!\Phi_i\! \right> \,.
\label{cdi4lg}
\end{align}
\end{subequations}
The seagull term is not present in the length gauge.
In the next section, we analyze these corrections 
in the light of gauge invariance.
We separate the corrections by their physical origin, and show more than 
the gauge invariance of the final result: namely, we 
are able to demonstrate that 
each physically distinguished correction is gauge invariant in itself. 

%
%
\section{GENERAL PROOF OF GAUGE INVARIANCE}
\label{genproof}

%
%

\subsection{Orientation}

First of all, let us remember that in all bound-state calculations, we actually
use a hybrid gauge transformation~\cite{ScBeBeSc1984} where we ignore the gauge
transformation of the wave function. The non-interacting relativistic
Hamiltonian, given in Eq.~\eqref{H4}, by definition is gauge invariant. Thus,
we only gauge transform the electron's transition current and the photon field
operator, or alternatively, we let the interaction Hamiltonian undergo a gauge
transformation. We show here that the full gauge invariance is obtained by
carefully considering the interplay of the relativistic corrections to the wavefunction, to the Hamiltonian and to the energies of the bound states (the
initial and the final states).

The whole problem becomes simpler when it is divided into three distinct parts,
the first of which is a generalized correction due to the relativistic
Hamiltonian, the second of which is a quadrupole correction, and the third is a
remaining correction (a further correction to the current), which
can be shown to vanish after the use of commutator relations. Gauge
invariance can be shown for each of these corrections separately, provided some
parts of the velocity gauge correction to the electron's transition
current~\eqref{delJvg} are identified as being generated by the relativistic
Hamiltonian~\eqref{H4}, and treated together with the correction to the
Hamiltonian. Here, the velocity-gauge expression appears to be more
complicated. The quadrupole correction, by contrast, looks a little more
involved in the length gauge. Gauge invariance with respect to the
velocity gauge can be shown provided we include a part of the seagull
term~\eqref{seagull} into the velocity-gauge expression for the quadrupole
term. It is then relatively easy to show that all remaining terms vanish
separately.

In the following, we discuss the general approach to the 
proof of gauge invariance in some detail. Further 
aspects are elucidated in Appendices~\ref{appa} and~\ref{appb}.

%
%
\subsection{Correction to the Hamiltonian}
\label{corrham}

Let us discuss first the general paradigm and 
start with the corrections induced by the relativistic
Hamiltonian~\eqref{H4}.
The gauge invariance for the leading-order term 
(the nonrelativistic result) can be traced to the 
formula~\eqref{gaugesim},
\begin{equation}\label{dhgauge}
\xi = - \omega_1 \, \omega_2 \, \zeta \,,
\end{equation}
where $\xi$ represents the velocity-gauge form and 
$\zeta$ represents the length-gauge form.

Both $\xi$ and $\zeta$ actually carry superscripts $ij$,
which we suppress here to leave the notation compact,
as already discussed.
Let us now suppose that the total velocity-gauge 
correction due to the relativistic Hamiltonian 
can be expressed as $\delta\xi_H$, 
and the corresponding length-gauge expression is
$\delta\zeta_H$.
The precise definition of 
$\delta\xi_H$ and $\delta\zeta_H$ will be discussed later.
We are able to show the following gauge invariance relation,
\begin{equation}
\label{master}
\delta \xi_H = - \omega_1 \, \omega_2 \, \delta \zeta_H - 
\delta \omega_{\rm max} \, \omega_1 \, \zeta \,,
\end{equation}
based on which we can prove the gauge invariance of the 
entire correction $\delta \Gamma_H$ due to the 
relativistic Hamiltonian,
\begin{align}
\delta \Gamma^\xi_H &= 2\frac{4\alpha^2}{9\pi} \!\!\!\!\!\!\!\! 
\int \limits_{0}^{E_{\Phi_i}-E_{\Phi_f}} \!\!\!\!\!\!\!\!\! 
d \omega_1 \omega_1 \omega_2 \xi \delta \xi_H + 
\frac{4\alpha^2}{9\pi} \delta \omega_{\rm max} \!\!\!\!\!\!\!\!\!\! 
\int \limits_{0}^{E_{\Phi_i}-E_{\Phi_f}} \!\!\!\!\!\!\!\!\! 
d \omega_1 \omega_1 \xi^2 \nonumber \\
&= 2 \frac{4\alpha^2}{9\pi} \!\!\!\!\!\!\!\! 
\int \limits_{0}^{E_{\Phi_i}-E_{\Phi_f}} \!\!\!\!\!\!\!\!\! 
d \omega_1 \omega_1 \omega_2 (-\omega_1 \omega_2 \zeta)
[-\omega_1 \omega_2 \delta \zeta_H - 
\delta \omega_{\rm max} \omega_1 \zeta] \nonumber \\
& \quad + \frac{4\alpha^2}{9\pi} 
\delta \omega_{\rm max} \!\!\!\!\!\!\!\!\!\! 
\int \limits_{0}^{E_{\Phi_i}-E_{\Phi_f}} \!\!\!\!\!\!\!\!\! 
d \omega_1 \omega_1^3 \omega_2^2 \zeta^2 \label{gaurel} \\
&= 2\frac{4\alpha^2}{9\pi} \!\!\!\!\!\!\!\! 
\int \limits_{0}^{E_{\Phi_i}-E_{\Phi_f}} \!\!\!\!\!\!\!\!\! 
d \omega_1 \omega_1^3 \omega_2^3 \zeta \delta \zeta_H + 
3\frac{4\alpha^2}{9\pi} \delta \omega_{\rm max} \!\!\!\!\!\!\!\!\!\! 
\int \limits_{0}^{E_{\Phi_i}-E_{\Phi_f}} \!\!\!\!\!\!\!\!\! 
d \omega_1 \omega_1^3 \omega_2^2 \zeta^2 \nonumber \\
&= \delta \Gamma^\zeta_H \,. \nonumber
\end{align}
Here, again, the superscript $\xi$ denotes the velocity gauge,
whereas $\zeta$ denotes the length gauge.
We are indeed able to show such a relation for all three terms given
in~\eqref{H4}, but only if we include in the definition of 
$\delta\xi_H$ specific
corrections to the electron's transition current. 
Our gauge-invariance relation can be illustrated as follows. 
The correction $\delta \xi_H$ contains the wavefunction correction 
in the velocity gauge, the Hamiltonian correction in
velocity gauge, the energy correction in velocity gauge, and the seagull
term in velocity gauge, 
as well as the current correction due to the current operator
$\delta J_H^i \equiv -\ii [ r^i, \delta H]$. By contrast, $\delta \zeta_H$ 
equals the sum of the wavefunction correction in length
gauge, the Hamiltonian correction in length gauge, and the energy
correction in length gauge. Note that 
the term $- \delta \omega_{\rm max} \, \omega_1 \, \zeta$ in
Eq.~\eqref{master} is related 
to the modified energy conservation condition, and that 
$\delta \omega_{\rm max}$ here is the correction to the 
transition frequency due to the relativistic Hamiltonian given in 
Eq.~\eqref{omegamax}.
Using this result, we are able to show that the 
total correction to the decay rate due to all three terms given in 
\eqref{H4} is gauge invariant. 

The current that we add in the velocity gauge is
\begin{align}
\label{defJH}
\delta J_H^i =& \; -i \left[ r^i, \delta H \right] =
-i \left[ r^i, - \frac{{\vec p}^{\,4}}{8 m^3} \right] 
-i \left[ r^i, \frac{\vec{L} \cdot \vec{\sigma}}{4m^2 r^3} \right]
\nonumber\\[2ex]
=& \;
- \frac{p^i \, {\vec p}^{\,2}}{2 m^3} 
-\frac{1}{4m^2} \frac{Z \alpha}{r^3} 
\left( \vec{r} \times \vec{\sigma} \right)^i \, .
\end{align}
The seagull term that we add in velocity gauge is 
due to a double commutator
\begin{align}
\label{defSH}
\delta S_H^{ij} =& \;  \left [ \left[r^i, \delta H \right],r^j \right] =
\left [ \left[r^i, -\frac{p^4}{8m^3} \right],r^j \right] 
\\[2ex]
=& \; \left[-i \, \frac{p^i p^2}{2m^3},r^j \right] 
= - \delta^{ij} \, \frac{p^2}{2m^3} - \frac{p^jp^i}{m^3} \,.
\nonumber
\end{align}
This term is part of the seagull Hamiltonian~\eqref{seagull}.
We are now in the position to give the precise definition of 
$\delta \xi_H$ and $\delta \zeta_H$, 
\begin{equation}
\label{defxiH}
\delta \xi_H = \sum_{i=1}^8 \delta \xi_i + 
\sum_{i=9}^{12} \delta \xi_i  \biggr|_{\delta J = \delta J_H} +
\delta \xi_{13} \biggr|_{\delta S = \delta S_H}
\end{equation}
and
\begin{equation}
\label{defzetaH}
\delta \zeta_H = \sum_{i=1}^8 \delta \zeta_i \,.
\end{equation}
For further details, see Appendix~\ref{appa}.

%
%

\subsection{Quadrupole (multipole) correction}
\label{corrmul}

The quadrupole correction is not associated with any 
correction to the bound-state energy or to the 
Schr\"{o}dinger Hamiltonian.
It can be treated separately and identified with a 
correction $\delta J_Q^i$ to the current in velocity gauge,
and with a correction $\delta I_Q^i$ in length gauge.
The velocity-gauge current is
\begin{align}
\delta J^i_Q =&\;
\frac{p^i}{m} \left( -\ii \vec{k} \cdot \vec{r} \right)
- \frac{1}{2} \frac{p^i}{m} (\vec{k} \cdot \vec{r})^2 
\nonumber\\[2ex]
\to &\;
- \frac{1}{2} \frac{p^i}{m} (\vec{k} \cdot \vec{r})^2 \label{JQ} \,.
\end{align}
We can ignore the first term because it vanishes
after angular algebra, for the first-order correction to the 
two-photon decay. This is unlike the $(Z\alpha)^2$ correction
to the Lamb shift, where this term contributes 
as a simultaneous perturbation to both currents, because
one and the same photon is being emitted. Here, 
two photons are being emitted, and angular averaging
occurs for both of them separately.

The quadrupole current in the length gauge is 
\begin{align}
\delta I^i_Q &= r^i \!
\left(- \frac{\mathrm{i}}{2} \vec{k} \cdot \vec{r} 
-\frac{1}{6} (\vec{k} \cdot \vec{r})^2 \right) +
\frac{1}{2m\omega } ( \vec{L} \times \vec{k} )^i 
\nonumber\\
& \quad - \frac{\ii}{6m\omega } 
\left[ ( \vec{L} \times \vec{k} )^i (\vec{k} \cdot \vec{r}) 
+ (\vec{k} \cdot \vec{r}) (\vec{L} \times \vec{k})^i \right]
\nonumber\\[2ex]
&\to r^i \!
\left( -\frac{1}{6} (\vec{k} \cdot \vec{r})^2 \right)
\nonumber\\
& \quad - \frac{\ii}{6m\omega } 
\left[ ( \vec{L} \times \vec{k} )^i (\vec{k} \cdot \vec{r}) 
+ (\vec{k} \cdot \vec{r}) (\vec{L} \times \vec{k})^i \right] \label{IQ}\,,
\end{align}
where in the last step we have
ignored the terms that vanish after angular integration.
We find that the quadrupole term is gauge invariant provided
we include, in the velocity-gauge expression, the seagull
contribution from the term 
\begin{equation}
\label{SQ}
\delta S_Q^{ij} = -\frac{1}{2m} (\vec{k} \cdot \vec{r})^2 \delta^{ij} \,.
\end{equation}
Now, the sum of $\delta S_Q^{ij}$ and $\delta S_H^{ij}$ is
the full higher-order seagull term
$\delta S^{ij}$ given in Eq.~\eqref{seagull}.

We denote the correction to the quadrupole matrix element 
in the velocity gauge by $\delta \xi_Q$ (it includes the 
seagull correction due to $\delta S^{ij}_Q$) and use 
$\delta \zeta_Q$ for the corresponding correction to the 
matrix element in the length gauge. We are able to 
show that 
\begin{align}
\delta \Gamma^\xi_Q &= 2\frac{4\alpha^2}{9\pi} \!\!\!\!\!\!\!\! 
\int \limits_{0}^{E_{\Phi_i}-E_{\Phi_f}} \!\!\!\!\!\!\!\!\! 
d \omega_1 \omega_1 \omega_2 \xi \delta \xi_Q \nonumber \\
&= 2 \frac{4\alpha^2}{9\pi} \!\!\!\!\!\!\!\! 
\int \limits_{0}^{E_{\Phi_i}-E_{\Phi_f}} \!\!\!\!\!\!\!\!\! 
d \omega_1 \omega_1 \omega_2 (-\omega_1 \omega_2 \zeta)
[-\omega_1 \omega_2 \delta \zeta_Q ] \label{gauquad} \\
&= 2\frac{4\alpha^2}{9\pi} \!\!\!\!\!\!\!\! 
\int \limits_{0}^{E_{\Phi_i}-E_{\Phi_f}} \!\!\!\!\!\!\!\!\! 
d \omega_1 \omega_1^3 \omega_2^3 \zeta \delta \zeta_Q 
= \delta \Gamma^\zeta_Q \,, \nonumber
\end{align}
proving the gauge invariance of the quadrupole 
correction. The precise definition of
$\delta \xi_Q$ and $\delta \zeta_Q$ reads as follows,
\begin{equation}
\label{defxiQ}
\delta \xi_Q = 
\sum_{i=9}^{12} \delta \xi_i  \biggr|_{\delta J = \delta J_Q} +
\delta \xi_{13} \biggr|_{\delta S = \delta S_Q}
\end{equation}
and
\begin{equation}
\label{defzetaQ}
\delta \zeta_Q = 
\sum_{i=9}^{12} \delta \zeta_i  \biggr|_{\delta I = \delta I_Q} \,.
\end{equation}
Further details are provided in Appendix~\ref{appb}.

%
%

\subsection{Remaining corrections}
\label{corrrem}

We have by now treated the correction due to the 
entire Hamiltonian~\eqref{H4}, the entire seagull
term~\eqref{seagull} and the quadrupole interaction.
The remaining terms are current corrections.
In the velocity gauge, these read
\begin{align}
\label{JR}
\delta J^i_R =& \;  \delta J^i - \delta J^i_H - \delta J^i_Q 
\nonumber\\[2ex]
= & \; -\frac{\mathrm{i}}{2m} 
\left( \vec{\sigma} \times \vec{k} \right)^i \! 
\left( 1\!-\! \mathrm{i}\vec{k} \cdot \vec{r} \right)\! 
-\frac{1}{4m^2} \frac{Z \alpha}{r^3} 
\left( \vec{r} \times \vec{\sigma} \right)^i .
\end{align}
Using commutator relations, it is possible to show that
\begin{align}
\label{42}
& \left< \Phi_f \left| p^i \frac{1}{H\! - \!E_{\Phi_i}\!+\!\omega} 
( \vec{\sigma} \times \vec{k})^i 
\right| \Phi_i \right> \nonumber \\
&+\left< \Phi_f \left| ( \vec{\sigma} \times \vec{k})^j 
\frac{1}{H\! -\!E_{\Phi_i}\!-\!\omega} p^j 
\right| \Phi_i \right> = 0 \,.
\end{align}
This relation
is valid for both $\vec k = \vec k_{1,2}$ if $\omega$ 
is changed according to Eqs.~\eqref{cdj1vg} to \eqref{cdj4vg}, 
and for an arbitrary
initial and final state. Thus, the contribution of the first term 
on the right-hand side of~\eqref{JR} vanishes.
Furthermore, we can replace
\begin{align}
& - \frac{1}{2m} \left( \vec{\sigma} \times \vec{k} \right)^i 
\left( \vec{k} \cdot \vec{r} \right) \to
-\frac{\ii \omega}{4m^2} \left( \vec{\sigma} \times \vec{p} \right)^i \,,
\nonumber\\[2ex]
& -\frac{1}{4m^2} \frac{Z \alpha}{r^3} 
\left( \vec{r} \times \vec{\sigma} \right)^i \to
\frac{\ii \omega}{4m^2} \left( \vec{\sigma} \times \vec{p} \right)^i \,,
\end{align}
when contracted with the photon propagator.
This relation is known from Lamb shift calculations
(see Ref.~\cite{Je2003dipl}). Therefore, the entire contribution
from the remaining corrections to the current vanishes in the 
velocity gauge.

In the length gauge, the remaining corrections to the 
current are given as
\begin{equation}
\delta I^i_R = \frac{1}{2m\omega} 
\left( \vec{\sigma} \times \vec{k} \right)^i 
\left( 1 - 
\mathrm{i}\vec{k} \cdot \vec{r} \right)  +\frac{\mathrm{i}\omega}{4m} 
\left( \vec{\sigma} \times \vec{r} \right)^i 
\,. \label{delIrm}
\end{equation}
The first term vanishes in view of Eq.~\eqref{42}.
The remaining terms also do not contribute to the corrections to
the decay rate. This follows from the relation
\begin{equation}
\frac{\mathrm{i}}{2m\omega} \left( \vec{\sigma} 
\times \vec{k} \right)^i \left( \vec{k} 
\cdot \vec{r} \right) \to  \frac{\mathrm{i}\omega}{4m} 
\left( \vec{\sigma} \times \vec{r} \right)^i \,.
\end{equation}
for the last two terms of
Eq.~\eqref{delIrm} when contracted with the
photon propagator. 
The precise definition of
$\delta \xi_R$ and $\delta \zeta_R$ reads as follows,
\begin{equation}
\label{defxiR}
\delta \xi_R = 
\sum_{i=9}^{12} \delta \xi_i  \biggr|_{\delta J = \delta J_R} 
\end{equation}
and
\begin{equation}
\label{defzetaR}
\delta \zeta_R = 
\sum_{i=9}^{12} \delta \zeta_i  \biggr|_{\delta I = \delta I_R} \,.
\end{equation}

%
%
\section{NUMERICAL CALCULATIONS}
\label{numerics}

%
%
\subsection{$\bm{2S}$--$\bm{1S}$ Decay}
\label{2s1s}

The phenomenologically most important two-photon decay process 
is the $2S$-$1S$ decay. Our gauge-invariant 
result for the correction
to the decay rate due to the relativistic Hamiltonian,
as discussed in Sec.~\ref{corrham}, reads
\begin{equation}
\label{GammaH}
\delta \Gamma_H = \Gamma_0 \, \left[ -0.5082 \, (Z\alpha)^2 \right]
\end{equation}
For the quadrupole correction, the gauge-invariant result is
(see Sec.~\ref{corrmul})
\begin{equation}
\label{GammaQ}
\delta \Gamma_Q = \Gamma_0 \, \left[ -0.1555 \, (Z\alpha)^2 \right]
\end{equation}
The remaining current corrections vanish, as discussed in 
Sec.~\ref{corrrem}, 
\begin{equation}
\delta \Gamma_R = 0 \,.
\end{equation}
The total result for the relativistic correction to the two-photon
decay rate thus reads
\begin{equation}
\label{finalres}
\delta \Gamma =
\delta \Gamma_H +
\delta \Gamma_Q +
\delta \Gamma_R =
\Gamma_0 \, \left[ -0.6636 \, (Z\alpha)^2 \right] \,.
\end{equation}
It is instructive to break down the corrections to the Hamiltonian 
further. Namely, according to Eq.~\eqref{H4}, we have the 
zitterbewegung (zb) term,
\begin{equation}
\label{Hzb}
\delta H_{\mathrm{zb}} = \frac{\pi Z \alpha}{2m} \delta^3 (\vec{r}) \,,
\end{equation}
the kinetic energy (ke) term,
\begin{equation}
\label{Hke}
\delta H_{\mathrm{ke}} = - \frac{p^4}{8m^3} \,,
\end{equation}
and the spin-orbit (LS) coupling
\begin{equation}
\label{HLS}
\delta H_{LS} = \frac{Z \alpha}{4m^2} \frac{\vec{L}
\cdot \vec{\sigma}}{r^3} \,.
\end{equation}
The corresponding results read,
for the $2S$-$1S$ decay,
\begin{subequations} 
\begin{align} 
\label{gamma2zb}
\delta \Gamma_{\mathrm{zb}} =& \;
\Gamma_0 \, \left[ -0.7577 \, (Z\alpha)^2 \right] \,, \\
\label{gamma2ke}
\delta \Gamma_{\mathrm{ke}} =& \;
\Gamma_0 \, \left[ 0.2495 \, (Z\alpha)^2 \right] \,, \\
\label{gamma2LS}
\delta \Gamma_{LS} =& \; 0 \,.
\end{align} 
\end{subequations} 
This concludes our discussion of the 
two-photon decay of the $2S$ state, and we can 
now proceed to calculate decays from higher excited states.

\begin{table}[tbh]
\centering
\caption{\label{table1}
Results for the $\gamma_2$ coefficient as 
defined in Eq.~\eqref{gamma23}. This coefficients
gives the relativistic corrections to the 
two-photon decay rate.}
\begin{tabular}{l@{\hspace{0.2in}}c@{\hspace{0.2in}}c}
\hline
\hline
\rule[-2mm]{0mm}{6mm}
 &
\multicolumn{1}{c}{$\left| \Phi_f \right>= \left| 1S_{1/2} \right>$} &
\multicolumn{1}{c}{$\left| \Phi_f \right>= \left| 2S_{1/2} \right>$} \\ 
\hline
\rule[-2mm]{0mm}{6mm}
$\left| \Phi_i \right>= \left| 2S_{1/2} \right>$ & $ -0.6636$ & - \\
\rule[-2mm]{0mm}{6mm}
$\left| \Phi_i \right>= \left| 3S_{1/2} \right>$ & $ -2.6637$ & $ -1.7038$ \\
\rule[-2mm]{0mm}{6mm}
$\left| \Phi_i \right>= \left| 4S_{1/2} \right>$ & $ -4.5192$ & $ -7.8530$ \\
\hline
\rule[-2mm]{0mm}{6mm}
$\left| \Phi_i \right>= \left| 3D_{3/2} \right>$ & $ -2.2978$ & $  7.8533$ \\
\rule[-2mm]{0mm}{6mm}
$\left| \Phi_i \right>= \left| 3D_{5/2} \right>$ & $ -1.0981$ & $-22.2671$ \\
\hline
\hline
\end{tabular}
\end{table}

%
%

\subsection{Higher Excited States}
\label{higher}

In principle, one might assume that in order
to calculate the relativistic correction to the 
two-photon decay from higher excited states,
only the initial and 
final state wavefunctions have 
to be changed accordingly. However, 
historically the generalization to higher 
excited states has proven to be 
problematic. For higher excited states,
the two-photon 
transition can take place not only through 
virtual intermediate states with an equal or higher 
energy than the initial state, but also 
through cascades via intermediates states with a lower 
energy. For the $3S$ initial state, a decay 
via the cascade $3S$-$2P$-$1S$ is possible. 
The allowed cascade
transitions cause singularities in the 
propagators. As we are interested in the 
total decay rate, we integrate over the 
propagators and thereby also over the 
singularities. These singularities are quadratic and thus 
{\rm a priori} not integrable.

Finally, after some
discussion~\cite{Fl1984,CrTaSaCh1986,FlScMi1988,%
ChSu2008,ChSu2009,Je2007,Je2008},
the conclusion has been reached that the two-photon
correction to the decay width of the initial state
can be obtained using an integration prescription where
the double poles are treated in a manner inspired 
by quantum electrodynamics, where the photon energy integration
contour extends infinitesimally into the complex 
plane~\cite{Mo1974a,Je2009}.
Note that the two-photon correction 
thus obtained is a further correction that
has to be added to the one-photon decay width that 
is otherwise responsible for the cascade transition.
Using this procedure, we were able 
to determine the relativistic and multipole 
corrections to the nonrelativistic decay rate 
for many higher excited states
which fulfill the same gauge relations as 
for the 2$S$-1$S$ transition.
Final results are given in Table~\ref{table1}.

%
%
\section{Leading Logarithmic QED Corrections}
\label{log}

The zitterbewegung term in the relativistic 
Hamiltonian, according to Eq.~\eqref{Hzb},
is given as
$\delta H_{\mathrm{zb}} = \pi Z \alpha \, 
\delta^3 (\vec{r})/(2 m)$.
The effective potential that gives the leading QED
radiative corrections is
\begin{equation}
\label{deltaH}
\delta H_{\mathrm{rad}} = 
\frac{4 \,\alpha}{3} \, (Z \alpha) \, 
\ln[(Z\alpha)^{-2}] \, \frac{\delta^3 (\vec{r})}{m^2} \,.
\end{equation}
This relation implies that the $\gamma_3$ coefficient can be obtained as
$8\,\gamma_{2,{\rm zb}}/3$ where $\gamma_{2,{\rm zb}}$ is the contribution to
$\gamma_2$ caused exclusively by the zitterbewegung term.  As this contains no
spin dependence, the $\gamma_3$ coefficient is spin independent.  For the
$2S$-$1S$ transition, e.g., we have according to Eq.~\eqref{gamma2zb}, the
relation $\gamma_3 = \frac{8}{3} (-0.7577) = -2.0205$.  Results for other
transitions are given in Table~\ref{table2}.  
The $\gamma_3$ coefficient becomes numerically rather large for $3D$-$2S$
transitions.  Note that the correction is the same for decay from $3D_{3/2}$ and
$3D_{5/2}$ because the potential~\eqref{deltaH} does not involve any
spin-dependent terms.

\begin{table}[tbh]
\centering
\caption{\label{table2}
Results for $\gamma_3$ as defined in 
Eq.~\eqref{gamma23}.}
\begin{tabular}{l@{\hspace{0.2in}}c@{\hspace{0.2in}}c}
\hline
\hline
\rule[-2mm]{0mm}{6mm}
 &
\multicolumn{1}{c}{$\left| \Phi_f \right>= \left| 1S_{1/2} \right>$} &
\multicolumn{1}{c}{$\left| \Phi_f \right>= \left| 2S_{1/2} \right>$} \\ 
\hline
\rule[-2mm]{0mm}{6mm}
$\left| \Phi_i \right>= \left| 2S_{1/2} \right>$ & $ -2.0203$ & - \\
\rule[-2mm]{0mm}{6mm}
$\left| \Phi_i \right>= \left| 3S_{1/2} \right>$ & $  9.6521$ & $ 16.0424$ \\
\rule[-2mm]{0mm}{6mm}
$\left| \Phi_i \right>= \left| 4S_{1/2} \right>$ & $ 20.7364$ & $ 61.7499$ \\
\hline
\rule[-2mm]{0mm}{6mm}
$\left| \Phi_i \right>= \left| 3D_{3/2} \right>$ & $ -5.4681$ & $144.3639$ \\
\rule[-2mm]{0mm}{6mm}
$\left| \Phi_i \right>= \left| 3D_{5/2} \right>$ & $ -5.4681$ & $144.3639$ \\
\hline
\hline
\end{tabular}
\end{table}

%
%
\section{CONCLUSIONS}
\label{Conclusions}

The precise treatment of the two-photon decay width 
in ionic hydrogenlike bound systems with low nuclear charge
numbers demands an evaluation of the relativistic and 
multipole correction of relative order $(Z\alpha)^2$, which 
is the leading correction to the classic result~\cite{GM1931}.
The leading logarithmic QED correction of relative 
order $\alpha\, (Z\alpha)^2 \, \ln[(Z\alpha)^{-2}]$
also needs to be determined. These corrections
can be parametrized according to Eq.~\eqref{gamma23} 
in terms of two coefficients $\gamma_2$ and $\gamma_3$
which are given in Tables~\ref{table1} and~\ref{table2}.

Of particular interest is the result
\begin{equation}
\label{gamma22S1S}
\gamma_2(2S-1S) = -0.6636
\end{equation}
for the $2S$-$1S$ decay.
This result [see Eq.~\eqref{finalres}]
is the sum of a correction due to the relativistic 
Hamiltonian [Eq.~\eqref{GammaH}] and a correction due to the 
quadrupole term [Eq.~\eqref{GammaQ}].
We also generalize our approach to the two-photon decay from higher
excited states (Tables~\ref{table1} and~\ref{table2}).
As usual in quantum electrodynamic calculations, 
the magnitude of the correction terms grows 
with the principal quantum number.
The decay from $D$ states is also treated,
and it is worthwhile noting that 
the spin-independent logarithmic correction
terms of relative order $Z \alpha^2 \, \ln(Z\alpha)$
turn out to be large in magnitude
(see Table~\ref{table2}).
Finally, as shown in Appendix~\ref{comp} below, 
a comparison of our results to those of 
a nonperturbative (in $Z\alpha$) calculation for the 
$3S$-$1S$ decay (Ref.~\cite{JeSu2008}) reveals that 
the term of relative order $(Z\alpha)^2$ can account for the bulk of the 
relativistic correction up to some rather high
nuclear charge numbers ($Z \lesssim 40$).

With our NRQED-inspired approach, we can 
uniquely identify the physical origin of the 
$(Z\alpha)^2$-correction terms to the 
two-photon decay width, as discussed in 
Secs.~\ref{corrham},~\ref{corrmul} and~\ref{corrrem},
and give their values separately.
It is sometimes worthwhile to use the effective 
nonrelativistic treatment of NRQED,
because it may yield information which 
could not be obtained by a fully 
relativistic treatment, regarding the breakdown of the 
corrections. Furthermore, 
the calculation of the full spectrum of 
the propagator can be greatly simplified using 
lattice methods~\cite{SaOe1989}, 
increasing the speed as well as the 
numerical stability of the evaluation,
which is especially important in the domain 
of low nuclear charge numbers.

Another aspect is that the proof of the gauge invariance, 
as carried out in full detail in Appendices~\ref{appa} and~\ref{appb},
turns out to be a surprisingly lengthy calculation.
We stress once more that 
the gauge invariance is shown to hold even if we 
ignore the gauge transformation of the wave function,
in the sense of the ``hybrid'' gauge transformation
developed in Refs.~\cite{ScBeBeSc1984,LaScSc1987}.

%
%
\section*{Acknowledgments}

This research was supported by the National Science Foundation
(Grant PHY--8555454) and by the Missouri Research Board.
B.J.W.~acknowledges support from the Deutsche Forschungsgemeinschaft
(contract Je285/4-1).
The authors acknowledge helpful conversations with K.~Pachucki.

\appendix

\begin{widetext}

%
%
\section{Gauge invariance of the Hamiltonian correction}
\label{appa}

We give further details regarding the 
gauge invariance of the seagull term. Useful 
general relations are
$p^i = \textrm{i} m \, [H-E+\omega, r^i]$
and 
$\omega_2 = E_{\Phi_i}-E_{\Phi_f}-\omega_1 $.
The term $\delta \xi_1$ can be transformed to
\begin{align}
\delta \xi_1 =& \left< \Phi_f \left| \frac{p^i}{m} \left( 
\frac{1}{H-E_{\Phi_i}+\omega_1} \right)^2 
\frac{p^j}{m} \right| \Phi_i \right> 
\langle \Phi_i | \delta H | \Phi_i \rangle =
- \omega_1 \omega_2 \left< \Phi_f \left| 
r^i \! \left(\! \frac{1}{H-E_{\Phi_i}+\omega_1} \!\right)^2 
\!\!r^j \right| \Phi_i \right> 
\langle \Phi_i | \delta H | \Phi_i \rangle 
\label{gauma1} 
\nonumber\\
& \; 
+ (\omega_2-\omega_1) \, \left< \Phi_f \left| r^i 
\frac{1}{H-E_{\Phi_i}+\omega_1} r^j \right| \Phi_i \right> 
\langle \Phi_i | \delta H | \Phi_i \rangle 
+ \left< \Phi_f \left| r^i r^j \right| \Phi_i \right> 
\langle \Phi_i | \delta H | \Phi_i \rangle \,.
\end{align}
An analogous relation also holds for $\delta \xi_2$,
\begin{align}
\delta \xi_2 =& 
\left< \Phi_f \left| \frac{p^i}{m} \left( 
\frac{1}{H-E_{\Phi_f}-\omega_1} 
\right)^2 \frac{p^j}{m} \right| \Phi_i \right> 
\langle \Phi_f | \delta H | \Phi_f \rangle =
- \omega_1 \omega_2 \left< \Phi_f \left| r^i \! 
\left( \! \frac{1}{H-E_{\Phi_f}-\omega_1} \! \right)^2 
\!\! r^j \right| \Phi_i \right> 
\langle \Phi_f | \delta H | \Phi_f \rangle 
\nonumber \\
& \quad 
+ (\omega_1-\omega_2) 
\left< \Phi_f \left| r^i \frac{1}{H-E_{\Phi_f}-\omega_1} 
r^j \right| \Phi_i \right> 
\langle \Phi_f | \delta H | \Phi_f \rangle 
+ \left< \Phi_f \left| r^i r^j \right| \Phi_i \right>
\langle \Phi_f | \delta H | \Phi_f \rangle  \,.
\end{align}
These relations are equal to those found 
in Ref.~\cite{Je2004rad} for a radiative 
correction potential. The relations for 
the correction to the wavefunctions are 
altered because we are considering a 
different Hamiltonian. 
Thus, $\delta \xi_3$ gives
\begin{align}
\delta \xi_3 =& \; 
\left< \Phi_f \left| \frac{p^i}{m} \frac{1}{H-E_{\Phi_i}+\omega_1} 
\frac{p^j}{m} \left( \frac{1}{E_{\Phi_i}-H} \right)' 
\delta H \right| \Phi_i \right> 
= - \omega_1 \omega_2 \! \left< \!\Phi_f \left| r^i 
\frac{1}{H\!-\!E_{\Phi_i}\!+\!\omega_1} r^j \! 
\left( \! \frac{1}{E_{\Phi_i}\!-\!H}\! \right)' 
\!\! \delta H \right|\! \Phi_i \right> 
\nonumber\\
& \; - \omega_2 \left< \Phi_f \left| r^i 
\frac{1}{H-E_{\Phi_i}+\omega_1} r^j \right| \Phi_i \right> 
\left< \Phi_i \left| \delta H \right| \Phi_i \right> 
+ \underbrace{\left< \Phi_f \left| r^i (H-E_{\Phi_i}+\omega_2) r^j 
\left( \frac{1}{E_{\Phi_i}-H} \right)' 
\delta H \right| \Phi_i \right>}_{\equiv T_3} 
\nonumber\\
& \; - \left< \Phi_f \left| r^i r^j \right| \Phi_i \right> 
\left< \Phi_i \left| \delta H \right| \Phi_i \right> 
+ \left< \Phi_f \left| r^i r^j \delta H \right| \Phi_i \right> 
+ \omega_2 \left< \Phi_f \left| r^i 
\frac{1}{H-E_{\Phi_i}+\omega_1} r^j \delta H \right| \Phi_i \right> \,.
\end{align}
For $\delta \xi_4$ this yields
\begin{align}
\delta \xi_4 =& \;
\left< \Phi_f \left| \frac{p^i}{m} \frac{1}{H-E_{\Phi_f}-\omega_1} 
\frac{p^j}{m} \left( \frac{1}{E_{\Phi_i}-H} \right)' 
\delta H \right| \Phi_i \right> 
= - \omega_1 \omega_2 \left< \!\Phi_f \left| r^i 
\frac{1}{H\!-\!E_{\Phi_f}\!-\!\omega_1} r^j  \! 
\left( \! \frac{1}{E_{\Phi_i}\!-\!H} \!\right)' \! 
\delta H \right| \!\Phi_i \right> \nonumber \\
& \; - \omega_1 \left< \Phi_f \left| r^i 
\frac{1}{H-E_{\Phi_f}-\omega_1} r^j \right| \Phi_i \right> 
\left< \Phi_i \left| \delta H \right| \Phi_i \right> 
+ \underbrace{\left< \Phi_f \left| r^i (H-E_{\Phi_f}-\omega_2) 
r^j \left( \frac{1}{E_{\Phi_i}-H} \right)' 
\delta H \right| \Phi_i \right>}_{\equiv T_{4}} \nonumber \\
& \; - \left< \Phi_f \left| r^i r^j \right| \Phi_i \right> 
\left< \Phi_i \left| \delta H \right| \Phi_i \right> 
+ \left< \Phi_f \left| r^i r^j \delta H \right| \Phi_i \right> 
+ \omega_1 \left< \Phi_f \left| r^i 
\frac{1}{H-E_{\Phi_f}-\omega_1} r^j \delta H \right| \Phi_i \right> \,.
\end{align}
For the correction $\delta \xi_5$ to the 
final-state wavefunction we get
\begin{align}
\delta \xi_5 =& \; \left< \Phi_f \left| \delta H \left( \frac{1}{E_{\Phi_f}-H} 
\right)' \frac{p^i}{m} \frac{1}{H-E_{\Phi_i}+\omega_1} 
\frac{p^j}{m} \right| \Phi_i \right> 
=- \omega_1 \omega_2 \left< \!\Phi_f \left| \delta H 
\! \left( \! \frac{1}{E_{\Phi_f}\!-\!H} \! \right)'\! r^i 
\frac{1}{H\!-\!E_{\Phi_i}\!+\!\omega_1} r^j 
\right|\! \Phi_i \right> 
\nonumber \\
& + \omega_1 
\left< \Phi_f \left| r^i 
\frac{1}{H-E_{\Phi_i}+\omega_1} r^j 
\right| \Phi_i \right> 
\left< \Phi_f \left| \delta H \right| \Phi_f \right> 
+ \underbrace{ \left< \Phi_f \left| \delta H \left( 
\frac{1}{E_{\Phi_f}-H} \right)' r^i (H-E_{\Phi_f}-\omega_1) 
r^j \right| \Phi_i \right> }_{\equiv T_5} 
\nonumber \\
& \quad - \left< \Phi_f \left| \delta H \right| \Phi_f \right> 
\left< \Phi_f \left| r^i r^j \right| \Phi_i \right> + 
\left< \Phi_f \left| \delta H r^i r^j 
\right| \Phi_i \right> 
- \omega_1 \left< \Phi_f \left| \delta H r^i 
\frac{1}{H-E_{\Phi_i}+\omega_1} r^j \right| \Phi_i \right> \,,
\end{align}
and for $\delta \xi_6$
\begin{align}
\delta \xi_6 =& \; \left< \Phi_f \left| \delta H \left( \frac{1}{E_{\Phi_f}-H} 
\right)' \frac{p^i}{m} \frac{1}{H-E_{\Phi_f}-\omega_1} 
\frac{p^j}{m} \right| \Phi_i \right> 
= - \omega_1 \omega_2 \left< \!\Phi_f \left| \delta H 
\! \left(\! \frac{1}{E_{\Phi_f}\!-\!H}\! \right)' \!\! 
r^i \frac{1}{H\!-\!E_{\Phi_f}\!-\!\omega_1} r^j 
\right|\! \Phi_i \right> 
\nonumber \\
& \quad + \omega_2 \left< \Phi_f \left| \delta H 
\right| \Phi_f \right> \left< \Phi_f \left| r^i 
\frac{1}{H-E_{\Phi_f}-\omega_1} r^j \right| \Phi_i \right> 
+ \underbrace{ \left< \Phi_f \left| \delta H \left( 
\frac{1}{E_{\Phi_f}-H} \right)' r^i (H-E_{\Phi_i}+\omega_1) 
r^j \right| \Phi_i \right> }_{\equiv T_6} \nonumber \\
& \quad - \left< \Phi_f \left| \delta H \right| \Phi_f \right> 
\left< \Phi_f \left| r^i r^j \right| \Phi_i \right> + 
\left< \Phi_f \left| \delta H r^i r^j \right| \Phi_i \right> 
- \omega_2 \left< \Phi_f \left| \delta H r^i 
\frac{1}{H-E_{\Phi_f}-\omega_1} r^j \right| \Phi_i \right> \,.
\end{align}
However, the corrections to the wavefunctions lead 
to some remainder terms which have to be analyzed 
separately. They can be transformed to give
\begin{align}
T_3 \! + \! T_4 =& \;
= \frac{1}{m} \left< \Phi_f \left| \left( 
\frac{1}{E_{\Phi_i}-H} \right)' \delta H 
\right| \Phi_i \right> \delta^{ij} 
 + \left< \Phi_f \left| r^i r^j \right| \Phi_i \right> 
\left< \Phi_i \left| \delta H \right| \Phi_i \right> - 
\left< \Phi_f \left| r^i r^j \delta H 
\right| \Phi_i \right> \,,
\\[2ex]
T_5 \! + \! T_6 =& \; 
= \frac{1}{m} \left< \Phi_f \left| \delta H \left( 
\frac{1}{E_{\Phi_f}-H} \right)' 
\right| \Phi_i \right> \delta^{ij} 
+ \left< \Phi_f \left| \delta H 
\right| \Phi_f \right> \left< \Phi_f \left| r^i r^j 
\right| \Phi_i \right> - \left< \Phi_f \left| \delta H r^i r^j 
\right| \Phi_i \right> \,. 
\end{align}
We observe the seagull terms $\delta \xi_{14}$ 
and $\delta \xi_{15}$ emerge and cancel, explicitly.
The other terms on the right-hand side 
will be treated separately, later.
The term $\delta \xi_7$ arising from the correction 
of the Hamiltonian can be brought into length gauge form
in the following way:
\begin{align}
\delta \xi_7 =& \; - \left< \Phi_f \left| \frac{p^i}{m} 
\frac{1}{H-E_{\Phi_i}+\omega_1} \delta H \frac{1}{H-E_{\Phi_i}+\omega_1} 
\frac{p^j}{m} \right| \Phi_i \right> 
= \omega_1 \omega_2 \left< \Phi_f \left| r^i 
\frac{1}{H\!-\!E_{\Phi_i}\!+\!\omega_1} \delta H 
\frac{1}{H\!-\!E_{\Phi_i}\!+\!\omega_1} r^j 
\right| \Phi_i \right> \,.
\nonumber\\[2ex]
& \; -\omega_2 \left< \Phi_f \left| r^i \frac{1}{H-E_{\Phi_i}+\omega_1} 
\delta H r^j \right| \Phi_i \right> 
\!+\! \omega_1 \! \left< \Phi_f \left| 
r^i \delta H \frac{1}{H\!-\!E_{\Phi_i}\!+\!\omega_1} 
r^j \right| \Phi_i \right> 
- \! \left< \Phi_f \left| r^i \delta H r^j \right| \Phi_i \right> \,.
\end{align}
Finally, for $\delta \xi_8$ we have
\begin{align}
\delta \xi_8 =& \;
- \left< \Phi_f \left| \frac{p^i}{m} \frac{1}{H-E_{\Phi_f}-\omega_1} 
\delta H \frac{1}{H-E_{\Phi_f}-\omega_1} \frac{p^j}{m} 
\right| \Phi_i \right> 
= \omega_1 \omega_2 \left< \Phi_f \left| r^i 
\frac{1}{H\!-\!E_{\Phi_f}\!-\!\omega_1} \delta H 
\frac{1}{H\!-\!E_{\Phi_f}\!-\!\omega_1} r^j 
\right| \Phi_i \right> 
\nonumber \\
& -\omega_1 \left< \Phi_f \left| r^i \frac{1}{H-E_{\Phi_f}-\omega_1} 
\delta H r^j \right| \Phi_i \right>
+\! \omega_2\! \left< \Phi_f \left| r^i \delta H 
\frac{1}{H\!-\!E_{\Phi_f}\!-\!\omega_1} r^j 
\right| \Phi_i\! \right> 
- \left< \Phi_f \left| r^i \delta H r^j \right| \Phi_i \right> \,.
\end{align}
Our intermediate result thus reads as follows,
\begin{align}
\label{hgauge1}
\sum_{i=1}^8 \delta \xi_i =&
-\omega_1 \omega_2 \sum_{i=1}^8 \delta \zeta_i
- \delta\omega_{\rm{max}} \omega_1 \zeta 
+ \omega_2 \left< \!\Phi_f \!\left| r^i \frac{1}{H-E_{\Phi_i}+\omega_1} 
[ r^j, \delta H ] \right| \Phi_i \right>
+ \omega_1 \left< \!\Phi_f \!\left| r^i \frac{1}{H-E_{\Phi_f}-\omega_1} 
[ r^j, \delta H ] \right| \Phi_i \right>
\nonumber\\[2ex]
& \; + \omega_2 \left< \Phi_f \left| [ r^i, \delta H ] 
\frac{1}{H-E_{\Phi_i}+\omega_1} r^j \right| \Phi_i \right>
+ \omega_1 \left< \Phi_f \left| [ r^i, \delta H ] 
\frac{1}{H-E_{\Phi_f}-\omega_1} r^j \right| \Phi_i \right> -
\left< \Phi_f \left| [ [ r^i, \delta H ] , r^j ] \right| \Phi_i \right> \,.
\end{align}
where $\delta \omega_{\rm{max}}$ is defined in Eq.~\eqref{omegamax}.
Fortunately, we can rewrite the terms with the $[ r^j, \delta H ]$
commutators further,
\begin{equation}
\label{hgauge2}
\begin{split}
& \omega_1\! \left< \Phi_f \left| [r^i,\delta H] 
\frac{1}{H\!-\!E_{\Phi_i}\!+\!\omega_1} r^j \right| \Phi_i \right> 
+ \omega_2 \left< \Phi_f \left| 
[r^i,\delta H] \frac{1}{H-E_{\Phi_f}-\omega_1} 
r^j \right| \Phi_i \right> \\
& \qquad + \omega_2 \left< \Phi_f \left| r^i 
\frac{1}{H-E_{\Phi_i}+\omega_1} [r^j,\delta H] 
\right| \Phi_i \right> 
+ \omega_1 \left< \Phi_f \left| r^i \frac{1}{H-E_{\Phi_f}-\omega_1} 
[r^j,\delta H] \right| \Phi_i \right> \\
&= -\left< \Phi_f \left| \delta J_H
\frac{1}{H-E_{\Phi_i}+\omega_1} \frac{p^j}{m} \right| \Phi_i \right> 
-  \left< \Phi_f \left| \delta J_H 
\frac{1}{H-E_{\Phi_f}-\omega_1} \frac{p^j}{m} \right| \Phi_i \right> \\
& \qquad - \left< \Phi_f \left| \frac{p^i}{m} 
\frac{1}{H-E_{\Phi_i}+\omega_1} \delta J_H \right| \Phi_i \right> 
- \left< \Phi_f \left| \frac{p^i}{m} 
\frac{1}{H-E_{\Phi_f}-\omega_1} \delta J_H \right| \Phi_i \right> 
+ 2 \left< \Phi_f \left| [[r^i,\delta H],r^j] \right| \Phi_i \right> \,.
\\
& = 
- \sum_{i=9}^{12} \delta \xi_i \biggr|_{\delta J = \delta J_H}
+ 2 \left< \Phi_f \left| [[r^i,\delta H],r^j] \right| \Phi_i \right> \,.
\end{split}
\end{equation}
The current $J_H = -\ii [ r^i, \delta H]$ is defined in Eq.~\eqref{defJH}.
Combining \eqref{hgauge1} and~\eqref{hgauge2}, we obtain the 
relation
\begin{equation}
\sum_{i=1}^8 \delta \xi_i + \sum_{i=9}^{12} \delta \xi_i 
\biggr|_{\delta J = \delta J_H}
= -\omega_1 \omega_2 \sum_{i=1}^8 \delta \zeta_i -
\delta\omega_{\rm{max}} \omega_1 \, \zeta
- \delta \xi_{13} \biggr|_{ \delta S = \delta S_H } \,.
\end{equation}
With the definitions~\eqref{defxiH} and~\eqref{defzetaH},
this leads directly to our gauge invariance relation~\eqref{master}.

%
%
\section{Gauge invariance of the quadrupole correction}
\label{appb}

For the proof of gauge invariance of the quadrupole 
correction it is more  convenient to start from the 
length gauge expression. As the quadrupole term is
a correction to the transition current,
only the terms $\delta \zeta_{9 \dots 12}$ 
are relevant. The length-gauge transition current 
$\delta I$ is [see Eq.~\eqref{IQ}]
\begin{equation}
\delta I^i_Q = r^i
\left( -\frac{1}{6} (\vec{k} \cdot \vec{r})^2 \right) 
+ \frac{1}{6m\omega } \left[ ( \vec{L} \times \vec{k} )^i 
(- \ii \vec{k} \cdot \vec{r}) 
+ (- \ii \vec{k} \cdot \vec{r}) (\vec{L} \times \vec{k})^i \right] \,.
\end{equation}
It is helpful to rewrite the second 
part of the transition current as
\begin{equation}
( \vec{L} \times \vec{k})^i (-\mathrm{i} \vec{k}\cdot \vec{r}) 
+ (-\mathrm{i} \vec{k}\cdot \vec{r}) ( \vec{L} \times \vec{k})^i 
= (\vec{k} \cdot \vec{r}) p^i (-\mathrm{i} \vec{k}\cdot \vec{r}) 
- r^i (\vec{k}\cdot \vec{p}) (-\mathrm{i} \vec{k}\cdot \vec{r}) 
+ (-\mathrm{i} \vec{k}\cdot \vec{r}) (\vec{k}\cdot \vec{r}) p^i 
- (-\mathrm{i} \vec{k}\cdot \vec{r}) r^i (\vec{k}\cdot \vec{p}) \,.
\end{equation}
Using this and the general relations from 
Appendix \ref{appa} we can transform the first 
term $\delta \zeta_9$ to give
\begin{align}
- \omega_1 \omega_2 \delta \zeta_9 
&= \omega_2 \frac{1}{6} \left< \Phi_f \left| r^i \frac{1}{H\!-\!E_{\Phi_i}\!+\!\omega_1} \! 
\left( \! \omega_1 (\vec{k}_1 \cdot \vec{r})^2 r^j \! 
+ \frac{\ii}{m} \left[ ( \vec{L} \times \vec{k}_1 )^j 
( \vec{k}_1 \cdot \vec{r}) 
+ ( \vec{k}_1 \cdot \vec{r}) (\vec{L} \times \vec{k}_1)^j \right] \right) \right| \Phi_i \right> \nonumber \\
&= \left< \Phi_f \left| \frac{p^i}{m} 
\frac{1}{H-E_{\Phi_i}+\omega_1} \left[- \frac{1}{2} 
\left(\vec{k}_1 \cdot \vec{r} \right) \right] 
\frac{p^j}{m} \right| \Phi_i \right> 
- \frac{1}{6} \omega_1 \left< \Phi_f 
\left| r^i (\vec{k}_1 \cdot \vec{r})^2 r^j 
\right| \Phi_i \right> \nonumber \\
& \quad + \frac{\ii}{6} \left< \Phi_f \left| \frac{p^i}{m} 
(\vec{k}_1 \!\cdot\! \vec{r})^2 r^j \right| \Phi_i \right> 
- \left(\frac{\ii}{6} k^l_1 k^m_1 \right) \left< \Phi_f \left| r^i r^l 
\frac{p^j}{m} r^m - r^i r^j \frac{p^l}{m} r^m + r^i r^m r^l 
\frac{p^j}{m} - r^i r^m r^j \frac{p^l}{m} \right| \Phi_i \right> \,.
\end{align}
For $\delta \zeta_{10}$ we obtain in an analogous manner
\begin{align}
- \omega_1 \omega_2 \delta \zeta_{10} &= \omega_1 
\frac{1}{6} \left< \Phi_f \left| r^i 
\frac{1}{H\!-\!E_{\Phi_f}\!-\!\omega_1} \! 
\left( \!\omega_2 (\vec{k}_2 \cdot \vec{r})^2 r^j \! 
+ \frac{\ii}{m} \left[ ( \vec{L} \times \vec{k}_2 )^j 
( \vec{k}_2 \cdot \vec{r}) 
+ ( \vec{k}_2 \cdot \vec{r}) (\vec{L} \times \vec{k}_2)^j 
\right] \right) \right| \Phi_i \right> \nonumber \\
&= \left< \Phi_f \left| \frac{p^i}{m} 
\frac{1}{H-E_{\Phi_f}-\omega_1} \left[- \frac{1}{2} 
\left(\vec{k}_2 \cdot \vec{r} \right) \right] 
\frac{p^j}{m} \right| \Phi_i \right> - \frac{1}{6} 
\omega_2 \left< \Phi_f \left| r^i (\vec{k}_2 \cdot \vec{r})^2 
r^j \right| \Phi_i \right> \nonumber \\
& \quad + \frac{\ii}{6} \left< \Phi_f \left| \frac{p^i}{m} 
(\vec{k}_2 \!\cdot\! \vec{r})^2 r^j \right| \Phi_i \right> 
- \left(\frac{\ii}{6} k^l_2 k^m_2 \right) \left< \Phi_f \left| r^i r^l 
\frac{p^j}{m} r^m - r^i r^j \frac{p^l}{m} r^m + r^i r^m r^l 
\frac{p^j}{m} - r^i r^m r^j \frac{p^l}{m} \right| \Phi_i \right> \,.
\end{align}
For the correction $\delta \zeta_{11}$ with the current 
acting on the left side this yields 
\begin{align}
- \omega_1 \omega_2 \delta \zeta_{11} &= \omega_1 \frac{1}{6} 
\left< \Phi_f \left| \left( \!\omega_2
(\vec{k}_2 \cdot \vec{r})^2 r^i \! + \frac{\ii}{m} 
\left[ ( \vec{L} \times \vec{k}_2 )^i ( \vec{k}_2 \cdot \vec{r}) 
+ ( \vec{k}_2 \cdot \vec{r}) (\vec{L} \times 
\vec{k}_2)^i \right] \right) \frac{1}{H\!-\!E_{\Phi_i}\!+\!\omega_1} 
r^j \right| \Phi_i \right> \nonumber \\
&= \left< \Phi_f \left| \frac{p^i}{m} \left[- \frac{1}{2} 
\left(\vec{k}_2 \cdot \vec{r} \right) \right] 
\frac{1}{H-E_{\Phi_i}+\omega_1} \frac{p^j}{m} \right| \Phi_i \right> 
+ \frac{1}{6} \omega_2 \left< \Phi_f \left| r^i 
(\vec{k}_2 \cdot \vec{r})^2 r^j \right| \Phi_i \right> \nonumber \\
& \quad - \frac{\ii}{6} \left< \Phi_f \left| r^i 
(\vec{k}_2 \!\cdot\! \vec{r})^2 \frac{p^j}{m} \right| \Phi_i \right> 
+ \left(\frac{\ii}{6} k^l_2 k^m_2 \right) \left< \Phi_f \left| r^l 
\frac{p^i}{m} r^m r^j - r^i \frac{p^l}{m} r^m r^j + r^m r^l 
\frac{p^i}{m} r^j - r^m r^i \frac{p^l}{m} r^j \right| \Phi_i \right> \,,
\end{align}
and finally for $\delta \zeta_{12}$,
\begin{align}
- \omega_1 \omega_2 \delta \zeta_{12} &= \omega_2 \frac{1}{6} 
\left< \Phi_f \left| \left( \!\omega_1 (\vec{k}_1 \cdot \vec{r})^2 
r^i \! + \frac{\ii}{m} \left[ ( \vec{L} \times \vec{k}_1 )^i 
( \vec{k}_1 \cdot \vec{r}) + ( \vec{k}_1 \cdot \vec{r}) 
(\vec{L} \times \vec{k}_1)^i \right] \right) 
\frac{1}{H\!-\!E_{\Phi_f}\!-\!\omega_1} r^j 
\right| \Phi_i \right> \nonumber \\
&= \left< \Phi_f \left| \frac{p^i}{m} \left[- \frac{1}{2} 
\left(\vec{k}_1 \cdot \vec{r} \right) \right] 
\frac{1}{H-E_{\Phi_f}-\omega_1} \frac{p^j}{m} \right| \Phi_i \right> 
+ \frac{1}{6} \omega_1 \left< \Phi_f \left| r^i 
(\vec{k}_1 \cdot \vec{r})^2 r^j \right| \Phi_i \right> \nonumber \\
& \quad - \frac{\ii}{6} \left< \Phi_f \left| r^i 
(\vec{k}_1 \!\cdot\! \vec{r})^2 \frac{p^j}{m} \right| \Phi_i \right> 
+ \left( \frac{\ii}{6} k^l_1 k^m_1 \right) \left< \Phi_f \left| r^l 
\frac{p^i}{m} r^m r^j - r^i \frac{p^l}{m} r^m r^j + r^m r^l 
\frac{p^i}{m} r^j - r^m r^i \frac{p^l}{m} r^j \right| \Phi_i \right> \,.
\end{align}
Combining these results, we get
\begin{align}
- \omega_1 \omega_2 \sum_{i=9}^{12} \delta \zeta 
\biggl|_{\delta I = \delta I_Q} \!\! &= \sum_{i=9}^{12} \delta \xi 
\biggl|_{\delta J = \delta J_Q} \!\!
- \left(\frac{\ii}{6} k^l_1 k^m_1\right) \left< \Phi_f \left| 
r^i r^l \frac{p^j}{m} r^m - r^i r^j \frac{p^l}{m} r^m 
+ r^i r^m r^l \frac{p^j}{m} - r^i r^m r^j \frac{p^l}{m} 
- \frac{p^i}{m} r^l r^m r^j \right| \Phi_i \right> \nonumber \\
& \quad - \left(\frac{\ii}{6} k^l_2 k^m_2 \right) \left< \Phi_f \left| 
r^i r^l \frac{p^j}{m} r^m - r^i r^j \frac{p^l}{m} r^m 
+ r^i r^m r^l \frac{p^j}{m} - r^i r^m r^j \frac{p^l}{m} 
- \frac{p^i}{m} r^l r^m r^j \right| \Phi_i \right> \nonumber \\
& \quad + 
\left(\frac{\ii}{6} k^l_2 k^m_2 \right) \left< \Phi_f \left| 
r^l \frac{p^i}{m} r^m r^j - r^i \frac{p^l}{m} r^m r^j 
+ r^m r^l \frac{p^i}{m} r^j - r^m r^i \frac{p^l}{m} r^j 
- r^i r^l r^m \frac{p^j}{m} \right| \Phi_i \right> \nonumber \\
& \quad + 
\left(\frac{\ii}{6} k^l_1 k^m_1 \right) \left< \Phi_f \left| 
r^l \frac{p^i}{m} r^m r^j - r^i \frac{p^l}{m} r^m r^j 
+ r^m r^l \frac{p^i}{m} r^j - r^m r^i \frac{p^l}{m} r^j 
- r^i r^l r^m \frac{p^j}{m} \right| \Phi_i \right> \,.
\end{align}
In order to simplify the resulting expression,
we now commute the momentum operators in 
the remainder terms to the right,
\begin{align}
- \omega_1 \omega_2 \sum_{i=9}^{12} \delta \zeta 
\biggl|_{\delta I = \delta I_Q} &= \sum_{i=9}^{12} 
\delta \xi \biggl|_{\delta J = \delta J_Q} 
+ \left< \Phi_f \left| \frac{1}{2m} \delta^{ij} 
\left( \vec{k}_1 \cdot \vec{r} \right)^2 
\right| \Phi_i \right> + \left< \Phi_f \left| \frac{1}{2m} 
\delta^{ij} \left( \vec{k}_2 \cdot \vec{r} \right)^2  
\right| \Phi_i \right> \,.
\end{align}
The last two terms can be identified as the negative 
of the quadrupole contribution to the 
higher-order seagull term as given in Eq.~\eqref{SQ},
summed over the two photon momenta $k_1$ and $k_2$.
Finally, this leads to the equality
\begin{align}
- \omega_1 \omega_2 \sum_{i=9}^{12} 
\delta \zeta \biggl|_{\delta I = \delta I_Q} &= 
\sum_{i=9}^{12} \delta \xi \biggl|_{\delta J = \delta J_Q} + 
\delta \xi_{13} \biggl|_{\delta S = \delta S_Q} \,,
\end{align}
which verifies the gauge invariance relation given in 
Eq.~\eqref{gauquad}.

\end{widetext}

%
%
\section{Comparison of Analytic and Numerical Results}
\label{comp}

We would like to compare our results for 
the analytic coefficients listed in 
Tables~\ref{table1} and~\ref{table2} to 
numerical data obtained for $2S$-$1S$
(see Ref.~\cite{GoDr1981}) and 
$3S$-$1S$ (see Ref.~\cite{JeSu2008}). 
The authors of~Ref.~\cite{GoDr1981} obtained
a fit to a convenient functional form in $Z\alpha$,
leading to an approximate formula valid across the 
whole range of nuclear charge numbers $Z$
[see Ref.~\cite{GoDr1981} and 
also Eq.~(4.16) of Ref.~\cite{BeKlSh1997}],
\begin{equation}
\label{prev}
\Gamma \approx \Gamma_0 \, 
\frac{1 + 3.9448 \, (Z\alpha)^2 - 2.040 \, (Z\alpha)^4}%
{1 + 4.6019 (Z\alpha)^2} \,.
\end{equation}
Upon re-expansion in $Z\alpha$, one may thus hope to
obtain an estimate for the correction of relative 
order $(Z\alpha)^2$.
Indeed, the estimate thus obtained,
$\gamma_2 \approx - 0.6571$, is in fair agreement
with the precise result~\eqref{gamma22S1S},
which reads $\gamma_2 = -0.6636$.

For the $3S$-$1S$ decay, we compare 
to a fully relativistic calculation carried
out in Ref.~\cite{JeSu2008}, where the relativistic effects have been
calculated for different values of $Z$. When using
our results for
$\gamma_2$,
one can determine the corrected decay rate for different values of $Z$.
For $Z=40$, our analytic results augmented by the 
relativistic correction of relative order 
$(Z\alpha)^2$, lead to a result of 
$\Gamma \approx 1.61 \, (Z=40)^6 \, {\rm rad}/{\rm s}$,
to be compared with the 
result $\Gamma = 1.60 \, (Z=40)^6 \, {\rm rad}/{\rm s}$
from Ref.~\cite{JeSu2008} for the 
$E1E1$ two-photon decay rate.

In general, there is quite a
subtle interplay of the fully relativistic calculations with
the Dirac--Coulomb propagator,
which have meanwhile been done
for a number of QED and other problems, and
the $Z\alpha$-expansion approach: numerically more
accurate results can be obtained with the former,
and these are relevant especially for highly charged
ions, but the physical origin of the relativistic
corrections is much more transparent within the
$Z\alpha$-expansion. 
Furthermore, the analytic calculations allow
for a systematic expansion in powers of $\alpha$
and $Z\alpha$, as demonstrated in Eq.~\eqref{gamma23}.

\end{document}